\documentclass[twocolumn, astrosymb, nofootinbib, times,tighten]{aastex631}%{emulateapj}
\usepackage{amsmath}
\usepackage{graphicx}
\usepackage{xcolor}
\usepackage{hyperref}
\usepackage{xfrac}
\usepackage{cleveref}
\usepackage{xcolor}
\usepackage{mathtools}
\usepackage[caption=false]{subfig}

\newcommand{\beq}{\begin{equation}}
\newcommand{\eeq}{\end{equation}}
\newcommand{\bea}{\begin{eqnarray}}
\newcommand{\eea}{\end{eqnarray}}

\definecolor{HWblue}{rgb}{0, 0, 1.0}
\definecolor{HWgreen}{rgb}{0, 0.45, 0}
\definecolor{HWred}{rgb}{1.0, 0, 0}
\definecolor{HWpink}{rgb}{1.0, 0.54, 0.4}
\definecolor{HWmagenta}{rgb}{0.7, 0.2, 0.64}
\usepackage[normalem]{ulem}

\newcommand{\rev}[1]{{\color{black}#1}}

\begin{document}

\title{High-redshift, small-scale tests of ultralight axion dark matter using Hubble and Webb galaxy UV luminosities}

\author{Harrison\ Winch\email{harrison.winch@mail.utoronto.ca}}
\affil{Department of Astronomy and Astrophysics, University of Toronto, 50 St. George Street, Toronto, ON M5S 3H4, Canada}

\author{Keir K. Rogers}
\affil{Dunlap Institute for Astronomy and Astrophysics, University of Toronto, 50 St. George Street, Toronto, ON M5S 3H4, Canada}

\author{Ren\'ee Hlo\v zek}
\affil{Department of Astronomy and Astrophysics, University of Toronto, 50 St. George Street, Toronto, ON M5S 3H4, Canada}
\affil{Dunlap Institute for Astronomy and Astrophysics, University of Toronto, 50 St. George Street, Toronto, ON M5S 3H4, Canada}

\author{David J. E. Marsh}
\affiliation{Theoretical Particle Physics and Cosmology, King’s College London, Strand, London, WC2R 2LS, United Kingdom}

\begin{abstract}
    We calculate the abundance of UV-bright galaxies in the presence of ultralight axion (ULA) dark matter (DM), finding that axions suppress their formation with a non-trivial dependence on redshift and luminosity. We set limits on axion DM using UV luminosity function (UVLF) data, excluding a single axion as all the DM for $m_\textrm{ax} < 10^{-21.6}$ eV and limiting axions with $-26 \leq \log( m_\textrm{ax}/\mathrm{eV}) \leq -23$ to be less than $22\%$ of the DM (both at $95\%$ credibility). These limits use UVLF measurements from 24,000 sources from the Hubble Space Telescope (HST) at redshifts $4 \leq z \leq 10$. We marginalize over a parametric model connecting halo mass and UV luminosity. Our results bridge a window in axion mass and fraction previously unconstrained by cosmological data, between large-scale CMB and galaxy clustering and the small-scale Lyman-$\alpha$ forest. These high-$z$ measurements provide a powerful consistency check of low-$z$ tests of axion DM, including the recent hint for a sub-dominant ULA DM fraction in Lyman-$\alpha$ forest data. We also consider a sample of 25 spectroscopically-confirmed high-$z$ galaxies from the James Webb Space Telescope (JWST), finding these data to be consistent with HST. Combining HST and JWST UVLF data does not improve our constraints beyond HST alone, but future JWST measurements have the potential to improve these results. We also find an excess of low-mass halos ($< 10^9 M_\odot$) at $z < 3$, which could be probed by sub-galactic structure probes (e.g., stellar streams, satellite galaxies and strong lensing).

\end{abstract}

%notes from Keir, March 18th
%fig 2, change minimum ly-a to z=2 DONE
%fig 4, att arrows
%add figure with HMF varying axion fraction, to show origins of the spoon DONE
%fig 11, run with HST+Planck best-fit
%gif 13, add note about -26 not being rules out entirely, just making statement about powterior volume
%table 1, add column with mean +- sigma, and best-fit
%add point in discussion about how we're ignoring all axion wave effects, assuming they will all be below the halo jeans scale, talk about axion simulations that could help with that
%add numbers or fractions for sub-chains for mass bin plot
%appendix B talk more about why we're neglecting model 1

\section{Introduction}

\label{sec:Intro} 

%introduce ULA DM

Ultralight axions and axion-like particles (with masses $ m_\mathrm{ax} \lesssim 10^{-18}$ eV) are well-motivated dark matter (DM) particle candidates. Ultralight axions arise generically from a variety of broken symmetries or compactified extra dimensions, manifesting as a largely non-interacting pseudo-Nambu-Goldstone boson that would behave as DM. These particles are therefore not only a compelling DM candidate, but also a powerful probe of string theory and high-energy physics \citep{Dine:1982ah,Preskill:1982cy,Abbott:1982af, Svrcek:2006yi,Conlon:2006tq,Arvanitaki:2009fg,Mehta:2021pwf, OHare:2024nmr}. Ultralight axions have a de Broglie wavelength that manifests on astrophysical scales, smoothing out structure below the axion Jeans scale $\lambda_J$, which depends on the axion particle mass like $\propto m_\mathrm{ax}^{-\sfrac{1}{2}}$ \citep{Amendola:2005ad, Duffy:2009ig, Arvanitaki:2009fg, Park2012,Hlozek:2014lca,  Marsh:2015xka, Hui:2016ltb, Lague:2021frh,OHare:2024nmr}. 
%This Jeans scale ($\lambda_J$) can be related to axion particle mass ($ m_\textrm{ax}$) by
%\begin{equation}
%    \frac{\lambda_J}{2.4 \text{ Mpc}} = h^{-1/2} \bigg(\frac{ m_\textrm{ax}}{10^{-25} \text{ eV}}\bigg)^{-1/2}.
%\end{equation}
%Thus, lower axion masses smooth structure on larger scales. 
%Throughout this paper we will refer to both ultralight axions and axion-like particles simply as `axions'.

The astrophysical ($\sim$ kpc to Gpc) scale of these axion wave features necessitates the use of cosmological observables in order to search for their effects. These searches often constrain both the axion particle mass and the axion DM fraction through their observational signatures \citep{Hlozek:2014lca,Marsh:2015xka, Kobayashi:2017jcf, Lague:2021frh, Rogers:2023ezo, Rogers:2023upm}. The discovery of a subdominant DM fraction of axions would have profound implications for fundamental physics. Indeed, many axion DM models motivated by string theory propose a range of axion particle masses existing simultaneously (sometimes called the `string axiverse', see \cite{Arvanitaki:2009fg, Marsh:2011gr, Gendler:2023kjt}). Some of these particles (such as the quantum chromodynamics (QCD) axion, see \cite{PecceiQuin:1977, PhysRevLett.40.223, Wilczek:1977pj, Berezhiani:1992rk,  DiLuzio:2020wdo}) would manifest as a cold DM (CDM) component when the axions have sufficiently high mass.
%Additionally, if one allows for tuning in the axion starting field angle, this can enhance cosmic power on certain scales \citep{Leong:2018opi, LinaresCedeno:2020dte, Arvanitaki:2019rax,  Winch:2023qzl}. Further discussion of these `extreme axions' is reserved for Sec. \ref{sec:Discussion}. 

\begin{figure*}
    \centering
    \includegraphics[trim={1.2cm 0.5cm 2.25cm 2.2cm},clip,width=0.85\linewidth]{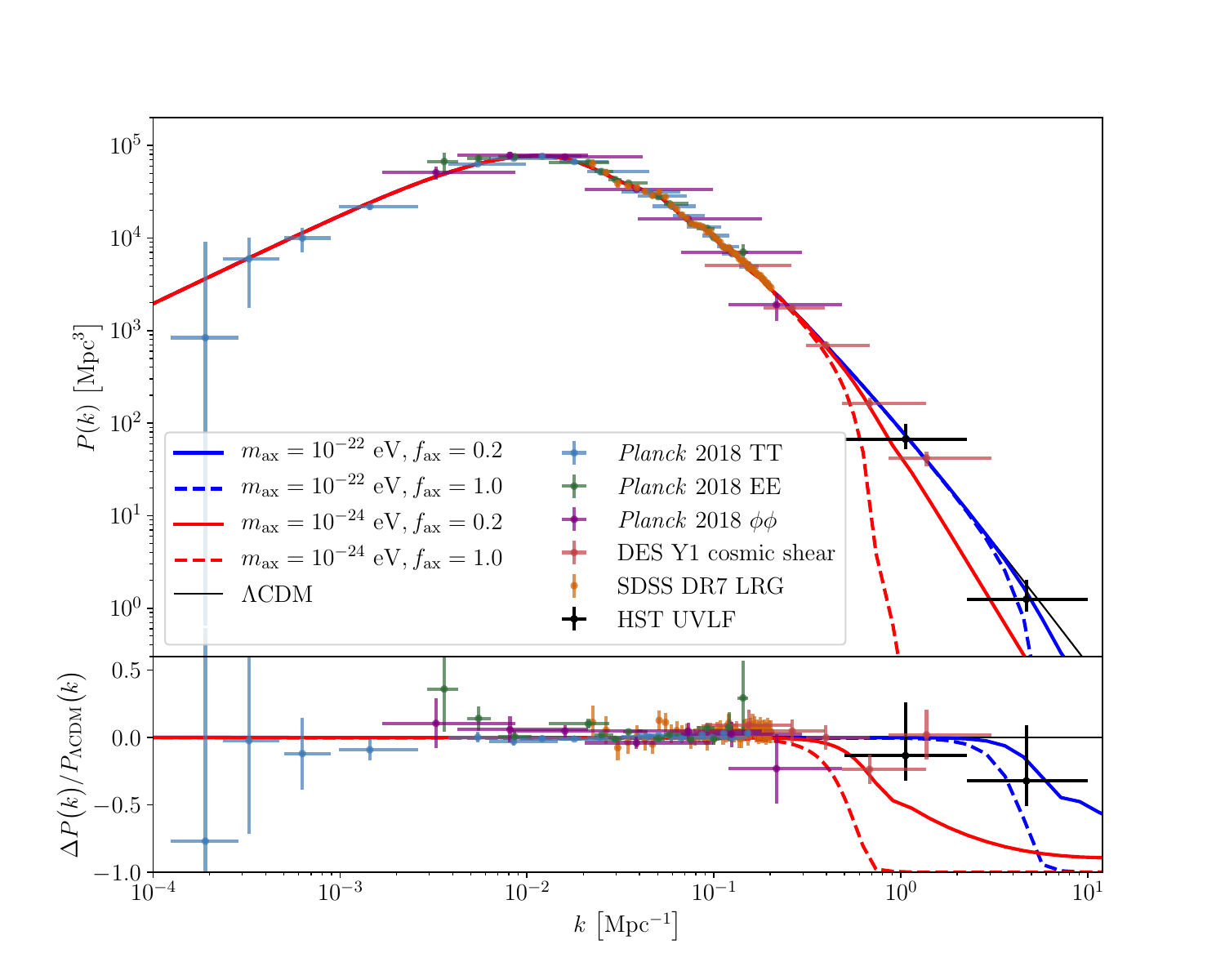}
    \caption{The linear matter power spectrum (MPS) as a function of wavenumber \(k\) for both cold dark matter (CDM, in black) and axion (in red and blue) cosmologies with different axion mass ($m_\mathrm{ax}$) and DM fraction ($f_\mathrm{ax}$). Blue, green, and magenta error bars represent constraints on the MPS from \textit{Planck} 2018 measurements of the cosmic microwave background (CMB) \citep{Planck:2018vyg} temperature TT, polarization EE and lensing \(\phi \phi\) angular power spectra, respectively. Red error bars are from galaxy weak lensing shear estimates made with the Dark Energy Survey \citep[DES;][]{DES:2017myr}, while orange error bars are from galaxy clustering estimates using the Sloan Digital Sky Survey \citep[SDSS;][]{SDSS_MPS}. Black error bars represent constraints on the MPS based on both \textit{Planck} CMB and Hubble Space Telescope (HST) measurements of the UVLF as computed in \cite{Sabti:2021unj}. We do not show Lyman-\(\alpha\) forest inference on the MPS as existing estimates do not account for the tension with \textit{Planck} cosmology in the tilt of the small-scale power (as discussed in \cite{Rogers:2023upm}). This plot was constructed using code from \cite{Sabti:2021unj} (based on code by \hyperlink{https://github.com/marius311/mpk_compilation}{Marius Millea}), as well as the axion Boltzmann code \texttt{axiCLASS} \citep{Poulin:2018dzj}. Previous iterations of this plot were presented in \cite{Tegmark:2002cy, ACT_MPS_Hlozek, Chabanier:2019eai}}
    \label{fig:power_spectrum}
\end{figure*}

\begin{figure}
    \centering
    \includegraphics[trim={0.3cm 1cm 1.2cm 2cm}, clip, width=\linewidth]{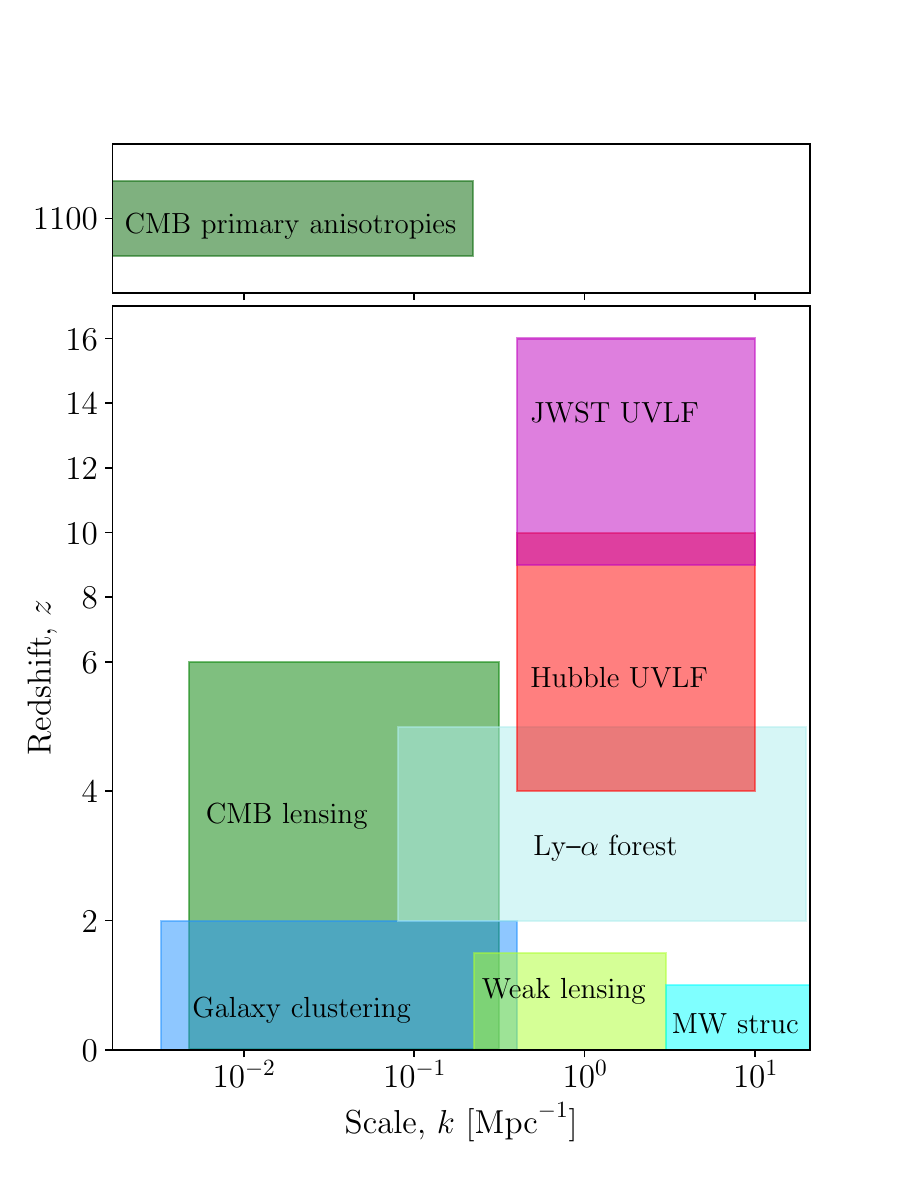}
    \caption{The approximate range of wavenumbers \(k\) and redshifts \(z\) probed by various astrophysical measurements capable of constraining axions and the nature of DM. Hubble and James Webb Space Telescope (JWST) measurements of the UVLF probe a unique range of small scales and high redshifts that are currently inaccessible otherwise.}
    \label{fig:kz_schematic}
\end{figure}

The effects of axion mass and density fraction on the linear matter power spectrum are shown in Fig.~\ref{fig:power_spectrum}. Fig.~\ref{fig:kz_schematic} illustrates the scales and redshifts probed by different cosmological observables. Large-scale measurements of the CMB and galaxy clustering have put tight constraints on the axion fraction for low-mass ($ m_\textrm{ax} \leq 10^{-25}$ eV) axions \citep{Hlozek:2014lca, Hlozek:2017zzf, Lague:2021frh, Rogers:2023ezo}. Small-scale measurements of the Lyman-$\alpha$ forest rule out higher-mass ($10^{-23} \textrm{ eV} \lesssim m_\textrm{ax} \lesssim 10^{-20}$ eV) axions at higher DM fractions \citep{Irsic:2017yje, Kobayashi:2017jcf, Rogers:2020ltq}. Analyses of the kinematics of dwarf galaxies have claimed to rule out axions with mass of $10^{-19}$ eV as 100\% of the DM \citep{Marsh:2018zyw,Dalal:2022rmp}. Other low-$z$ astrophysical probes of axions include galaxy weak lensing \citep{Dentler:2021zij}, galaxy strong lensing \citep{Shevchuk:2023ccb}, galaxy rotation curves \citep{Bar:2021kti} and supermassive black holes \citep{DeLaurentis:2022nrv}. Future 21 cm measurements have the potential to detect signatures of axion DM at high redshift \citep{Tegmark:2008au, HERA:2021noe, Hotinli:2021vxg, Flitter:2022pzf, Liu:2022iyy}. A joint analysis of CMB and Lyman-$\alpha$ forest measurements from the Extended Baryon Oscillation Spectroscopic Survey \citep[eBOSS;][]{eBOSS} finds that a non-zero axion density (\(m_\mathrm{ax} \sim 10^{-25}\,\mathrm{eV}\)) alleviates tension in measurement of the small-scale power \citep{Rogers:2023upm}, while respecting existing limits. Axions (\(m_\mathrm{ax} \sim 10^{-25}\,\mathrm{eV}\)) are also found to address the \(S_8\) cosmological parameter discrepancy \citep{Rogers:2023ezo}.

However, there remains a substantial gap of unconstrained axion masses, around $10^{-25} \textrm{ eV} \lesssim m_\textrm{ax} \lesssim 10^{-23}$ eV, which have evaded current cosmological constraints. Claimed constraints in this mass range from strong lensing \citep{Shevchuk:2023ccb} and galaxy rotation curves \citep{Bar:2021kti} depend on modeling the complex astrophysics of the soliton core in dense galactic environments, making it difficult to probe low axion fractions, which motivates the use of complementary cosmological probes to further probe this gap. In addition, all other axion probes use either low ($z \lesssim 5$) or recombination ($z \sim 1100$) redshifts, with no powerful probes during the redshifts of early structure formation ($4 \lesssim z \lesssim 16$). Any claimed evidence of axion DM as a resolution to low-$z$ tensions \cite[e.g.,][]{Rogers:2023upm, Rogers:2023ezo} needs to be corroborated by high-$z$ consistency checks that can probe similarly small scales. 
%However, these Lyman-$\alpha$ forest estimates often rely on complex hydrodynamical modeling of small-scale baryonic physics, which is both computationally intensive and potentially introduces astrophysical systematic uncertainties. As such, it is prudent to seek independent probes of small-scale power, in order to corroborate (or perhaps surpass) constraints on axion DM imposed by current methods such as the Ly-$\alpha$ forest.

%introduce UVLF, measured using HST and JWST

In this paper, we investigate the use of the galaxy UV luminosity function (UVLF) as an independent probe of ultralight axions, capable of probing a novel range of scales and redshifts unexplored by observations (as seen in Fig.~\ref{fig:kz_schematic}). The UVLF, $\Phi_\mathrm{UV}(M_\mathrm{UV}, z)$, is the number of UV sources (galaxies) per unit volume per unit UV magnitude $M_\mathrm{UV}$ at redshift $z$. This function depends on the astrophysical model of star formation within the galaxy which contributes to the UV luminosity through the number of bright young stars, as well as depending on the halo mass function (HMF), which describes the number density of halos of a given mass. Since the de Broglie wavelength of ultralight axions prevents them from clustering into halos below a certain scale, this will impact the HMF, which will in turn impact the UVLF \citep{Bozek:2014uqa,Schive:2015kza,Corasaniti:2016epp}. Thus, the UVLF has the potential to probe axion physics on small scales, beyond the reach of more established large-scale structure observations. We use the UVLF likelihood package \texttt{GALLUMI} \citep{Sabti:2021unj}, which computes the UVLF using the formalism we describe in Sections \ref{sec:HMF} and \ref{sec:UVLF}. \cite{Song:2016, Sabti:2021unj, Sabti:2021xvh} already demonstrated the power of the high-\(z\) UVLF in testing the standard cosmological model. We present here the first use of this modeling of the UVLF in testing a concrete example of beyond Standard Model physics.
%, and with complementary systematics and a novel redshift range compared to current astrophysical or Ly-$\alpha$ forest experiments. The potential power of these experiments is illustrated in Fig.~\ref{fig:kz_schematic}, which illustrates how the UVLF probes a range of scales and redshifts that are not probed by other cosmological methods. %We can see in Fig.~\ref{fig:power_spectrum} that axion physics manifests primarily on scales too small to be probed by the CMB on its own, but directly in the range of scales probed by Hubble estimates of the UVLF, motivating its use as a powerful tool for axion detection.

\begin{figure}
    \noindent
    \includegraphics[trim={0.2cm 0.4cm 0.2cm 0.25cm }, clip,width=\linewidth]{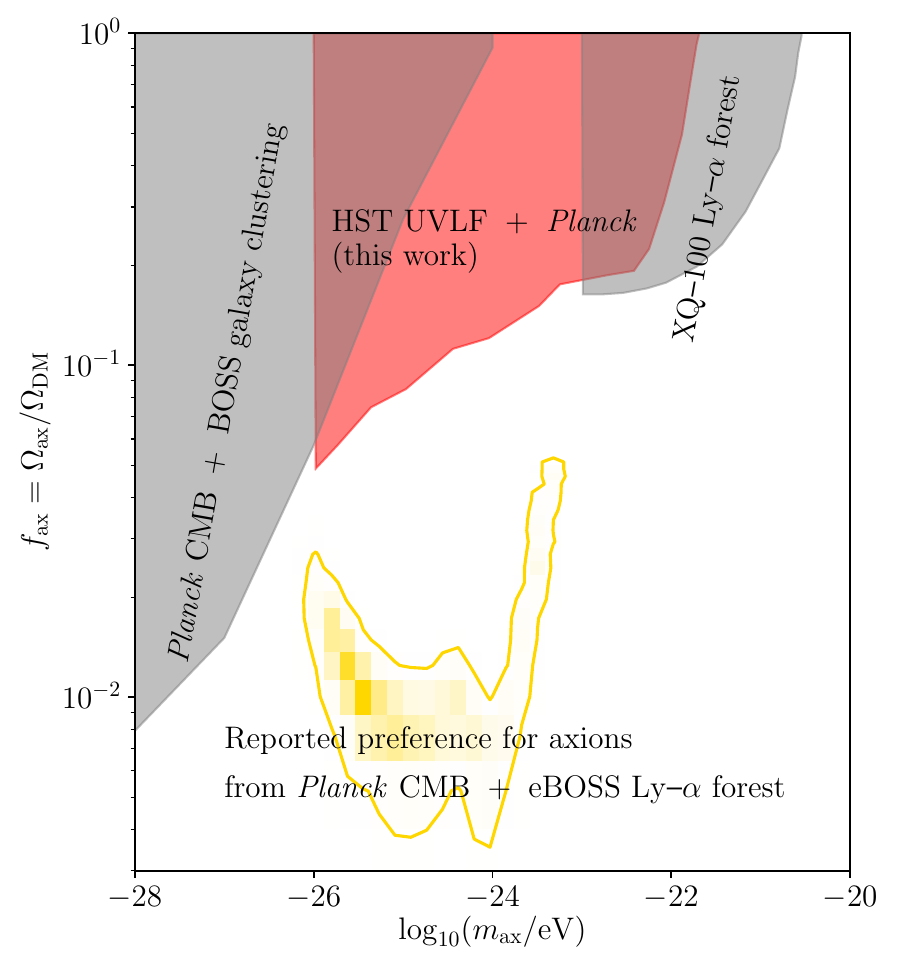}
    \caption{$2\sigma$ constraints on axion mass and DM fraction from cosmological probes (astrophysical constraints are neglected due to their complex dependence on soliton modeling). The right shaded region represents constraints using the Lyman-$\alpha$ forest computed in \cite{Kobayashi:2017jcf}, with data from the XQ-100 survey \citep{Lopez_XQ100} and the MIKE and HIRES spectrographs, which only considered axions down to a mass of $10^{-23}$ eV. The left shaded region represents joint constraints from \textit{Planck} CMB and BOSS galaxy clustering, as computed in \cite{Rogers:2023ezo}. This work (in red, also shown in Fig.~\ref{fig:joint_massfrac}) crucially fills a gap between these two other methods. The gold contours represent the reported preference for axion DM presented in \cite{Rogers:2023upm}, computed using both \textit{Planck} CMB and eBOSS Ly-$\alpha$ forest data, which is consistent with our UVLF limits.}
    \label{fig:joint_massfrac_compare}
\end{figure}

%\textcolor{red}{Added this bit comparing to previous works. Should also say something quantitative in discussion section.} 
Previous studies constraining axions with the UVLF include \cite{Bozek:2014uqa,Schive:2015kza,Corasaniti:2016epp,Menci:2017nsr,Leung:2018evj,Ni:2019qfa}. \cite{Schive:2015kza,Corasaniti:2016epp,Ni:2019qfa} use \(N\)-body simulations to compute the observables and were therefore unable to perform a detailed statistical analysis being limited by the number of simulations they could calculate. \cite{Bozek:2014uqa,Menci:2017nsr,Leung:2018evj} use similar semi-analytic methods to those we use here, but did not perform a full statistical analysis of constraints on mass and fraction. \rev{These studies use older HST datasets that do not extend to $z=10$, unlike our work.} These previous works find consistent results with one another, validating the semi-analytic methods. In the present work, we perform, for the first time, a complete statistical analysis combining CMB and UVLF data (see Fig.~\ref{fig:joint_massfrac_compare} for a summary of our main result).

The HST has observed over 24,000 UV sources between different surveys at $4 \leq z \leq 10$ \citep{HST_UVLF, 2022ApJ...940...55B}. 
%These measurements can be translated into estimates of the matter power spectrum on small scales, as shown in Fig.~\ref{fig:power_spectrum}, using a model of the relation between UV luminosity and halo mass as discussed above and in Section \ref{sec:UVLF}. 
These UVLF measurements have recently been augmented by groundbreaking new results from JWST, which can observe UV sources at much higher redshifts, up to $z \sim 16$ \citep{Bisigello2016, JWST_UVLF}. We find that while the JWST data do constrain the axion fraction on their own, the quality of the HST data still provides tighter constraints on axion physics.

Some early analyses of JWST observations have suggested that its estimates of the \(z > 10\) UVLF may be substantially higher than the $\Lambda$CDM expectations based on the \(z < 10\) HST estimates \citep{Bouwens2023, Chemerynska2023}. If true, this overabundance of structure at high redshifts could imply tight constraints on the allowed axion fraction, given that axions would decrease this UVLF (though this is a mass-dependent statement, as shown in Figure~\ref{fig:UVLF_mass}). 

However, there are significant uncertainties in star and galaxy formation at these high redshifts that need to be marginalized over, and so a full statistical analysis including a robust model of these uncertainties is essential to draw conclusions regarding the nature of DM. In addition, the discrepency appears to be higher in photometric JWST samples, where low-$z$ interlopers may impact the distribution of UV magnitudes.
%introduce sections of paper

%show lyman alpha k=20, z=5-6

%We calculate the impact of mixed axion and cold DM models on both the HMF and UVLF, before calculating constraints on the axion mass and fraction from current HST UVLF data. 
In Section \ref{sec:HMF}, we introduce our method for computing the HMF and present the impacts of mixed axion models on the HMF, while Section \ref{sec:UVLF} does the same for the UVLF. In Section \ref{sec:MCMC}, we present the results of a statistical comparison with data, using measurements of both HST UVLF and \textit{Planck} CMB. In Section \ref{sec:Discussion}, we compare these constraints to those from other probes and discuss the prospects for future improvements. We conclude in Section \ref{sec:Conclusions}.

%\section{Methods}
%\label{sec:Methods}

\section{The Halo Mass Function}
\label{sec:HMF}

The halo mass function (HMF) is the number density of dark matter halos of total mass $M_h$. This function, which varies with redshift, can be calculated for both $\Lambda$CDM and mixed axion models. It is a key component for calculating the UVLF, our primary observable, which is discussed in Section \ref{sec:UVLF}. Through this section and Sec.~\ref{sec:UVLF}, we will consider three possible impacts of axions on the UVLF, which are listed below:
\begin{itemize}
    \item \textbf{Cosmological impact of axions} refers to the impact of axions on the linear matter power spectrum, which goes into the computation of the HMF. We find that this mechanism has the dominant effect on the UVLF.
    \item \textbf{Axion halo pressure} refers to the effect of axions preventing the creation of halos below the halo Jeans limit due to their quantum pressure effects.
    \item \textbf{Axion astrophysics} refers to the impact of axions on the baryonic sector for halos below the Jeans limit. Halos below this limit will lack an axion component, due to axion halo pressure, and thus will have a greater ratio of baryons to DM. This will lead to increased star formation relative to total halo mass, which impacts the UVLF as described in Section \ref{sec:axions_UVLF}.
\end{itemize}

%introduce the three impacts, and what we're calling them.
%differentiate between cutoff mass and critical mass

\subsection{Computing the HMF}

The HMF can be computed using the method described in \cite{Jenkins:2000bv, Cooray:2002dia,  Sabti:2021xvh, Vogt:2022bwy}. The HMF is defined as
\begin{equation}
    n(M_h,z) \equiv \frac{1}{M_h} \frac{d \tilde{n}}{d \ln M_h},
\end{equation}
where $M_h$ is halo mass and $\tilde{n}$ is halo number density. Following the ellipsoidal collapse model of halo formation, we can write the HMF as
\begin{equation}
    n(M_h,z) = \frac{1}{2} \frac{\bar{\rho}(z)}{M_h^2} f(\nu) \bigg|\frac{d\ln\sigma^2}{d\ln M_h}\bigg|,
\end{equation}
where $\bar{\rho}(z)$ is the average total matter density at the relevant redshift and $\sigma^2$ is the variance of linear fluctuations, which is described in more detail below. The multiplicity function $f(\nu)$ is chosen to be the Sheth-Tormen fitting function \citep{Sheth:1999mn}:
\begin{equation}
\label{eq:ShethTormen}
   f(\nu) = A \sqrt{\frac{2}{\pi}}\sqrt{q}\nu (1 + (\sqrt{q}\nu)^{-2p})e^{-\frac{q\nu^2}{2}},
\end{equation}
where $\nu = \delta_{\textrm{crit}}/\sigma(M_h,z)$, with $\delta_{\textrm{crit}}$ being the critical linear density threshold for halo collapse.\footnote{Other forms of the mass function have been proposed, such as in \cite{Reed:2006rw}. \cite{Sabti:2021unj} found that UVLF constraints on cosmology are largely independent of the choice of mass function. We compare these choices in Appendix \ref{sec:AppC} and find that axions impact the different mass functions in nearly identical ways, meaning that our final constraints are largely independent of the choice of mass function.} The fitting parameters $A=0.3222$, $p=0.3$, $q = 0.707$ and $\delta_\textrm{crit} = 1.686$ in order to match simulations described in \cite{Sheth:1999mn}, following what was done in \cite{Sabti:2021xvh, Vogt:2022bwy}. We discuss the HMF modeling further in Section \ref{sec:Discussion}.

$\sigma^2(M_\mathrm{h},z)$ is the variance of linear fluctuations at redshift $z$ using a spherical real-space top hat filter $\hat{W}(M_h,R_h)$ with a radius $R_h$ such that the average enclosed mass is equal to $M_h$.\footnote{The use of this window function overestimates the number of DM halos in models with a high-\(k\) cutoff in the matter power spectrum \citep{Schneider:2013ria}, implying our constraints may be conservative. See, e.g., \cite{Du:2023jxh} and references cited therein for discussion of alternative $k$-space window functions. In the following analysis, the uncertainty in the low mass HMF caused by the window function is accounted for indirectly in our treatment of ``axion astrophysics'', which is discussed in Sec.~\ref{sec:axions_HMF}.} We use the Fourier transform of $\hat{W}(M_h,R_h)$, which we denote $W(M_h,k)$. The variance
\begin{equation}
    \sigma^2(M_h,z) = \int \frac{d^3k}{(2 \pi)^3} W^2(M_h,k) P^\mathrm{L}(k,z),
\end{equation}
where $k$ is the physical wavenumber and $P^\mathrm{L}(k,z)$ is the linear matter power spectrum at redshift $z$, which we compute using a cosmological axion Boltzmann code. In this work, we use the \texttt{axiCLASS} code \citep{Poulin:2018dzj, Smith:2019ihp}.

It follows that the final component of the HMF expression:
\begin{equation}
    \frac{d\ln\sigma^2}{d\ln M_h} = \frac{3}{\sigma^2 R_h^4 \pi^2} \int_0^{\infty} dk \frac{P^\mathrm{L}(k)}{k^2} {I}(k,R_h),
\end{equation}
where
\begin{multline}
   {I}(k,R_h) = (\sin(kR_h) - kR_h \cos(kR_h))\times \\
   \bigg[\sin(kR_h) \bigg(1 - \frac{3}{(kR_h)^2} \bigg) + \frac{3}{kR_h}\cos(kR_h) \bigg],
\end{multline}
which is derived from the derivative of $W(M_h,k)$.

%\subsection{Impact of axions on the HMF}

When the cosmological model contains both axions and CDM, the computation of the HMF will be impacted by the presence of axions. Galaxy halos are being formed from a mixture of axions and CDM \citep{Lague:2023wes}. 
As such, these halos will use the combined axion-and-CDM values of $P^\mathrm{L}(k,z)$, $\sigma(M_h,z)$, and $\bar{\rho}(z)$. Since axions alter the form of $P^\mathrm{L}(k,z)$ (and thus $\sigma(M_h,z)$), this will impact the shape of the HMF, even for large halos that may be above the axion Jeans scale.

\begin{figure}
    \centering
    \includegraphics[trim={0.4cm 0.9cm 1cm 1cm }, clip, width=\linewidth]{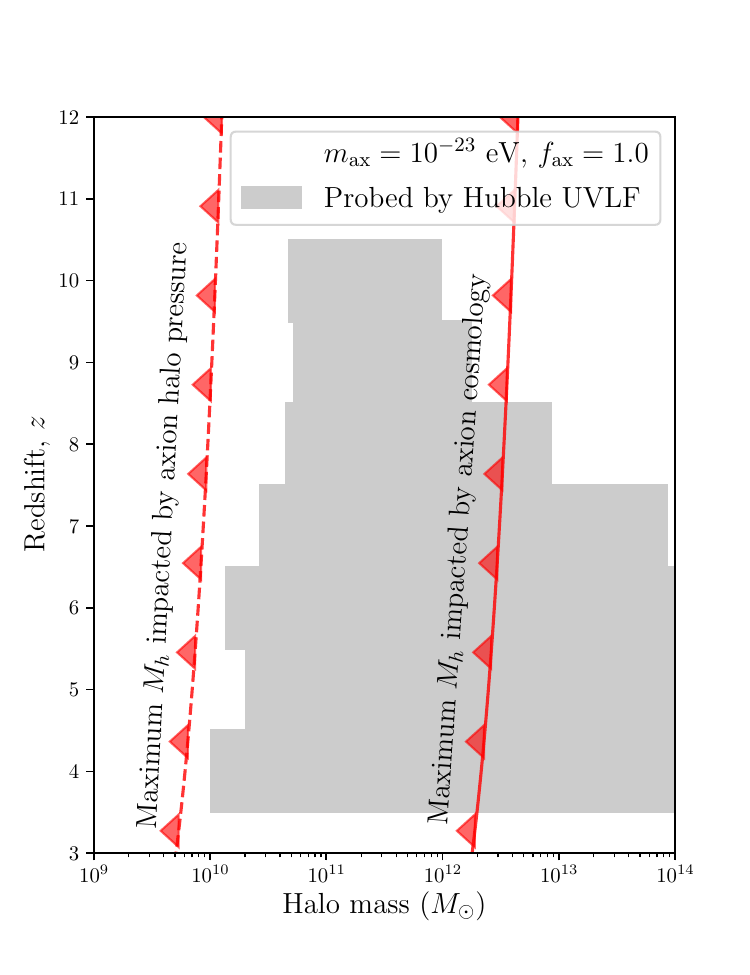}
    \caption{The redshifts and halo masses probed by HST measurements of the UVLF (grey boxes), along with the masses impacted by axion halo pressure and cosmology at $m_\mathrm{ax} = 10^{-23}$ eV (red lines). The halo masses impacted by axion astrophysics are always much smaller than those impacted by axion cosmology. This includes both the prevention of axion halo formation, as discussed in Sec.~\ref{sec:HMF}, as well as the indirect effects on the star formation rate below the Jeans scale, which is discussed in Sec.~\ref{sec:UVLF}.}
    \label{fig:Hubble_range}
\end{figure}

In addition to the cosmological impact through $P^\mathrm{L}(k,z)$, axion physics can impede the formation of halos of size equal to or smaller than the axion Jeans scale. This is due to the macroscopic de Broglie wavelength of ultralight axions preventing axions from clustering into small halos. In this work, we refer to this effect as ``axion halo pressure'' to distinguish its effects from the cosmological impact on the HMF, or the astrophysical impacts of axions on the UVLF (discussed in Sec~\ref{sec:axions_UVLF}). The axion Jeans wavenumber
\begin{equation}
    k_J \approx \sqrt{m_\mathrm{ax} H},
\end{equation}
where $m_\mathrm{ax}$ is the axion mass and $H$ is the Hubble parameter \citep{Hlozek:2014lca}. \cite{Marsh:2013ywa} and \cite{Vogt:2022bwy} compute a critical halo mass below which the average virial radius is below the axion Jeans scale, indicating that axion halo pressure would play a non-negligible role in halo formation. Such a critical halo mass
\begin{equation}
\label{eq:jeans_mass}
    M_{\mathrm{crit}} = \frac{4}{3}\pi \bigg(\frac{\lambda_J}{2}\bigg)^3 \rho_\mathrm{m},
\end{equation}
where $\lambda_J$ is the Jeans scale, related to the Jeans wavenumber by $\lambda_J/2 = \pi/k_J$. 

\begin{figure*}
    \centering
    \includegraphics[trim={1cm 1.2cm 2cm 1.5cm }, clip,width=\linewidth]{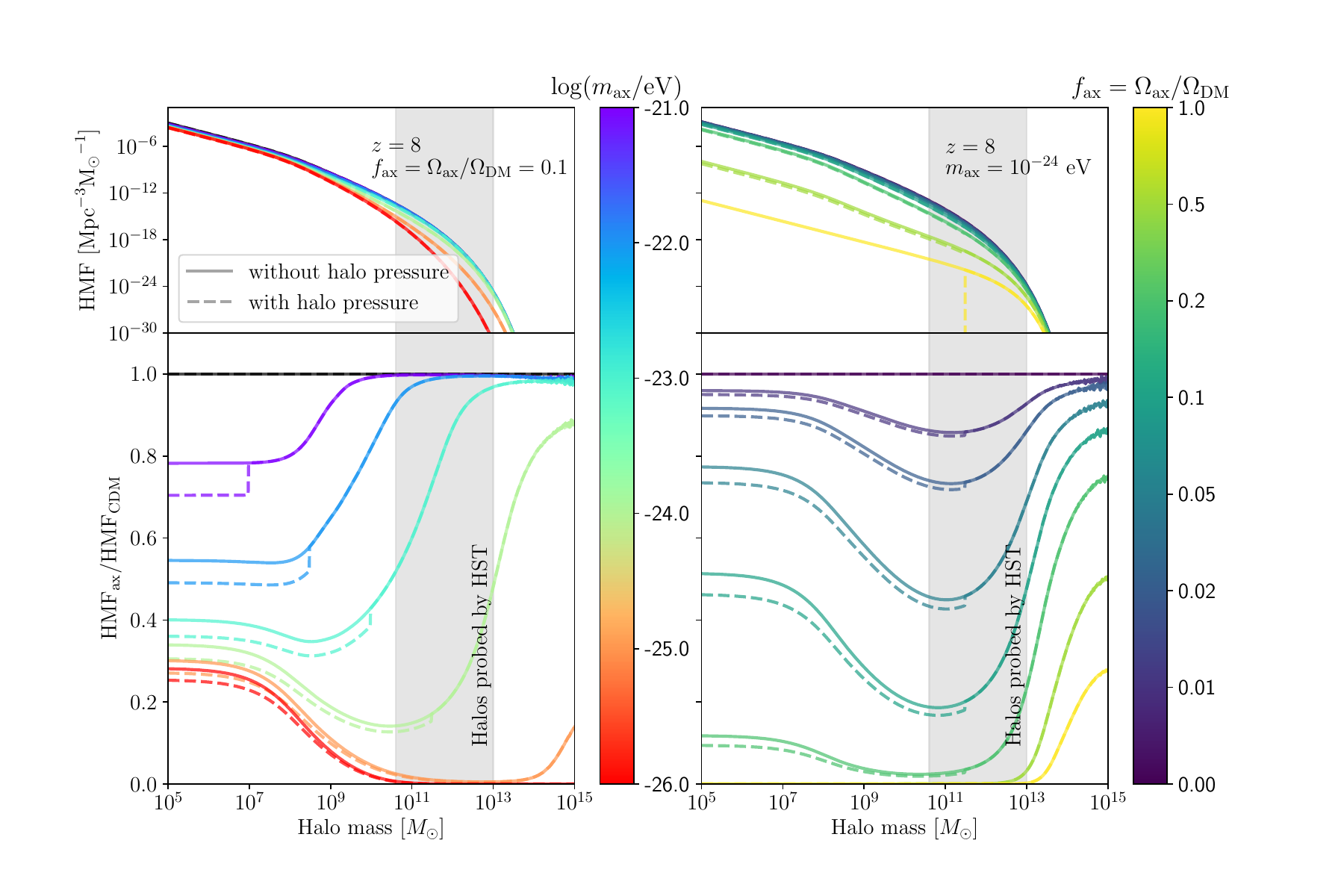}
    \caption{\rev{The left panels show the impact of axions on the HMF with varying mass at fixed axion fraction $f_\mathrm{ax} = \frac{\Omega_\mathrm{ax}}{\Omega_\mathrm{DM}} = 0.1$, while the right panels show the impact of varying axion fraction at fixed mass $m_\mathrm{ax} = 10^{-24}$ eV, all at $z=8$}. The top panels show the HMF, while the lower panels show the ratio of the HMF relative to a pure $\Lambda$CDM cosmology. Lower mass axions result in a stronger suppression of the HMF, along with a higher maximum impacted halo mass. The scale of axion fractions in the right panels is not linear. Higher axion fractions result in stronger suppressions of the HMF; for intermediate axion fractions the suppression is less pronounced at lower halo masses, as a result of delayed structure formation due to the suppression of initial small-scale fluctuations (see Fig.~\ref{fig:HMF_As} for an illustration of this process). We also plot the impact of axion halo pressure suppressing halo formation below the critical Jeans scale, shown with dashed lines. These corrections are negligible compared to the cosmological impact of axions on the HMF for the range of halo masses probed by HST. This is true for all axion masses and DM fractions that we consider.}
    \label{fig:HMF_axions}
\end{figure*}

The critical halo mass is always much smaller than the halo masses already impacted by the linear axion power spectrum. For many axion particle masses, the critical halo mass is too small even to be probed with the HST UVLF. Fig.~\ref{fig:Hubble_range} shows the range of halo masses probed by the HST UVLF at different redshifts (based on the UVLF model presented in Sec.~\ref{sec:UVLF}), compared to the maximum halos impacted by both the halo pressure and linear cosmology of axion DM with $m_\mathrm{ax} = 10^{-23}$ eV. The maximum halo mass impacted by axion halo pressure is not only smaller than the minimum halo probed by HST, but is also at least two orders of magnitude smaller than the maximum halo mass impacted by the axion linear cosmology. Both of these halo mass limits vary with axion mass in the same way, meaning that axion cosmology always impacts a substantially larger range of halo masses than the axion astrophysics \citep[see Fig. 2 in ][]{Marsh:2013ywa}. In this work, we ignore these axion halo pressure effects as negligible and consider only the cosmological impacts of axions on the HMF via $P^\mathrm{L}(k,z)$. This is consistent with the results of past works, including \cite{Dentler:2021zij}, which found that axion constraints from galaxy weak lensing using the Dark Energy Survey \citep{DES:2017myr} were not sensitive to the critical halo mass. We illustrate the negligible role of the axion halo pressure in Fig.~\ref{fig:HMF_axions}.

%We can model these effects by using the CDM-only values for $P^\mathrm{L}(k)$, $\sigma(M,z)$, and $\bar{\rho}(z)$ in the calculations of the HMF below this mass cutoff. This is a rough approximation - however, we show in Fig.~\ref{fig:HMF_m25_z} that these astrophysical effects are subdominant and occur on much smaller scales than the cosmological impacts on the HMF and UVLF.

%\begin{figure}
%    \centering
%    \includegraphics[trim={0cm 1cm 1cm 1cm }, clip,width=\linewidth]{plot_HMF_mass.pdf}
%    \caption{As Fig.~\ref{fig:HMF_axions} but for varying axion mass and \(f_\mathrm{ax} = 0.1\), \(z = 8\). Lower mass axions result in a stronger suppression of the HMF, along with a higher maximum impacted halo mass. This suppression appears to saturate at low halo masses.}
%    \label{fig:HMF_axions}
%\end{figure}

\subsection{Impact of axions on the HMF}
\label{sec:axions_HMF}

Considering only the cosmological impacts of axions on the HMF, there is a suppression in the number of halos below a certain mass. This suppression physically arises since, below their macroscopically-sized de Broglie wavelength, axions suppress the amplitude of density fluctuations that form halos. Fig.~\ref{fig:HMF_axions} (right-hand side) shows the impact of axions with a mass of $m_\mathrm{ax} = 10^{-24}$ eV on the HMF at $z=8$ for different axion fractions, showing how a higher axion fraction causes an increased suppression. Fig.~\ref{fig:HMF_axions} illustrates how allowing just 10\% of the DM to be in the form of axions can result in a significant suppression in the number of DM halos at that redshift (\(\sim 80\%\)). We consider $z=8$ as representative of the redshifts probed by the UVLF (see Figs.~\ref{fig:kz_schematic}, \ref{fig:Hubble_range} and ~\ref{fig:UVLF_fiducial_z}).

\begin{figure}
    \centering
    \includegraphics[trim={0.0cm 2.1cm 2cm 2.4cm }, clip,width=\linewidth]{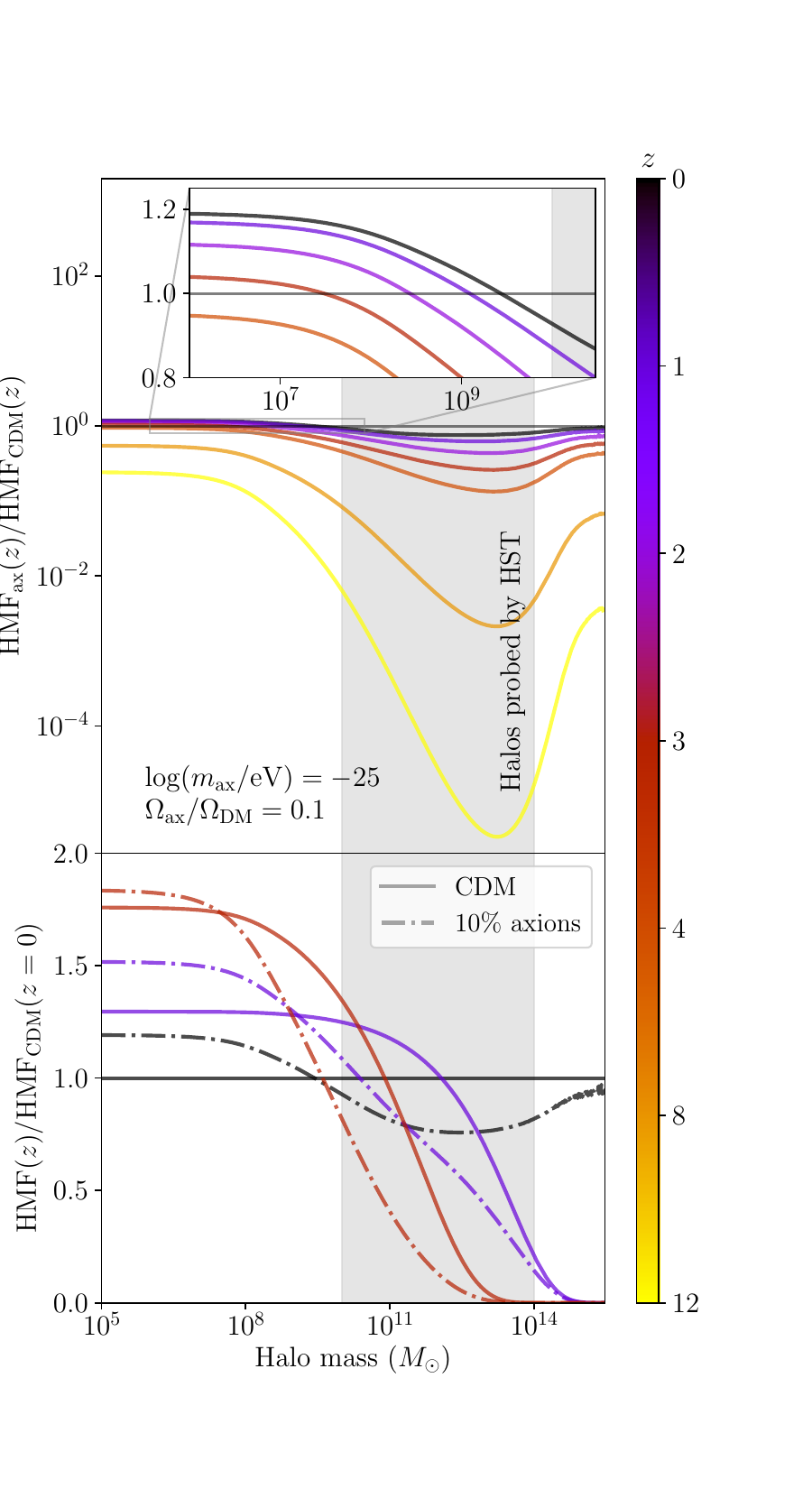}
    \caption{As Fig.~\ref{fig:HMF_axions}, but for varying redshift, \(m_\mathrm{ax} = 10^{-25}\,\mathrm{eV}\) and \(f_\mathrm{ax} = 0.1\). The upper plot shows the mixed-axion HMF relative to the $\Lambda$CDM HMF at the same redshift, with the inset focusing on low redshifts (between $z=0$ and $z=4$) where the HMF enhancement occurs. The lower plot shows a $\Lambda$CDM HMF (in solid) and a 10\%, $m_\mathrm{ax} = 10^{-25} \mathrm{\ eV}$ mixed-axion HMF (in dashed) at $z = 3,\ 1,\ \mathrm{and\ }0$, all relative to the $\Lambda$CDM HMF at $z=0$, in order to show how the excess low-mass HMF in the mixed axion model results from the delay in the growth of intermediate-mass halos. The grey band indicates halo masses probed by the HST UVLF across all redshifts, although the range for a particular redshift might be narrower than indicated (see Fig.~\ref{fig:Hubble_range} for an illustration of typical halo masses probed at a certain redshift).}
    \label{fig:HMF_m25_z}
    %add subplot showing comparison to CDM z=0
    %add line in caption explaining redshift range of HST sensitivities
\end{figure}

The left-hand side of Fig.~\ref{fig:HMF_axions} shows that higher mass axions (of order $m_\textrm{ax} \approx 10^{-21}$ eV \rev{as 10\% of the DM}) result in a suppression of around 20\% for halos lighter than $\sim 10^9 M_\odot$. However, lower mass axions (of order $m_\textrm{ax} \approx 10^{-24}$ eV \rev{as 10\% of the DM}) result in a significant suppression of halos with mass between $10^8$ and $10^{14} M_\odot$, with the suppression of halo formation reaching as much as 80\% compared to the $\Lambda$CDM model. The cutoff halo mass is roughly inversely proportional to the axion mass, as shown in \cite{Marsh:2013ywa}, since heavier axions have smaller wavelengths. However, the suppression for the lowest axion masses is less severe at low halo masses, plateauing at a suppression of less than 50\% for halo masses below $10^7 M_\odot$ for axion masses below $10^{-23}$ eV. This plateauing is due to the suppression of primordial structure on small scales causing a delay in hierarchical growth, resulting in more residual low-mass halos at late times that have not yet had a chance to merge into larger objects. We discuss this effect further below. This plateau is unlikely to be observed with the UVLF, where most galaxy halos that contribute to the Hubble and Webb UVLF have masses greater than $10^{10} M_\odot$ (as seen in Fig.~\ref{fig:Hubble_range} and by the grey bars in Figs.~\ref{fig:HMF_axions} and \ref{fig:HMF_m25_z}). We discuss the potential observability of this signature using low-$z$ probes of galaxy substructure in Sec.~\ref{sec:Discussion}.

Figure \ref{fig:HMF_axions} also shows the impact of axion halo pressure with dashed lines, which prevents the formation of axion halos below the critical Jeans halo mass (given in Eq.~\eqref{eq:jeans_mass}). The critical halo mass depends on the axion mass, where higher-mass axions have a shorter Jeans scale and thus suppress the formation of smaller halos. The amplitude of this suppression depends on the axion fraction, where models with higher axion DM fraction exhibit a greater suppression of halo formation on small scales. In both the left- and right-hand sides of Fig.~\ref{fig:HMF_axions}, the impact of axion halo pressure is always subdominant to the impact of axion cosmology, particularly in the range of halo masses probed by HST (in the light grey bar). Therefore, we safely neglect the effects of axion pressure on halo formation when computing the UVLF. Further, we explicitly check in Section \ref{sec:UVLF} the effect on the UVLF from axion halo pressure combined with ``axion astrophysics.''

\begin{figure}
    \centering
    \includegraphics[trim={0cm 0.2cm 1cm 1.0cm }, clip,width=\linewidth]{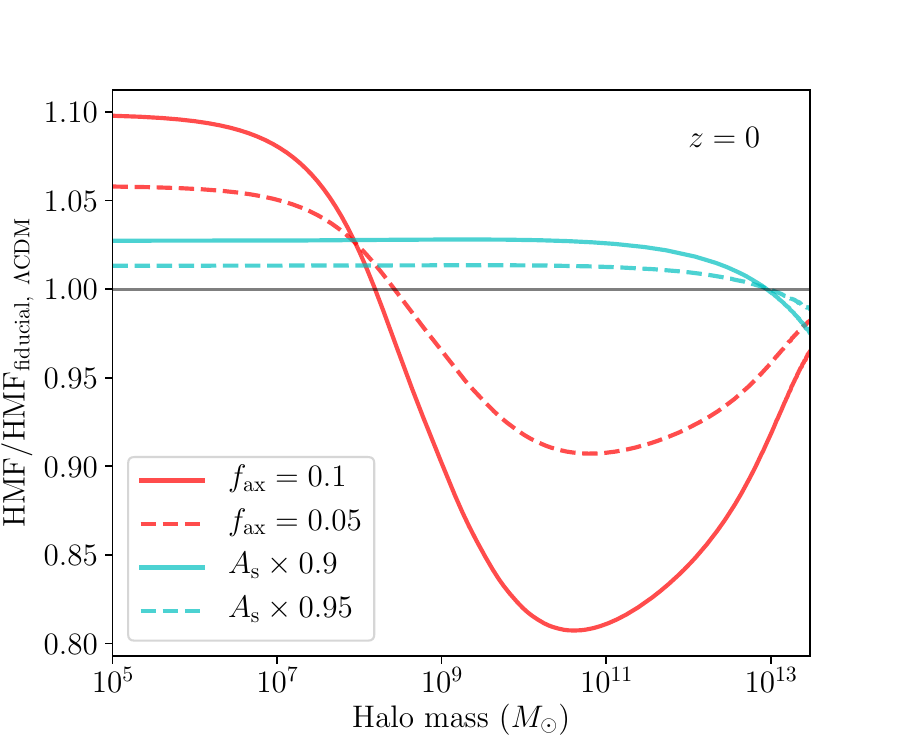}
    \caption{A comparison of the effect of axion fraction and primordial scalar amplitude $A_\mathrm{s}$ on the HMF at $z=0$. All lines are the ratio of the HMF with respect to a pure-CDM cosmology with a fiducial value of \(A_\mathrm{s} = 2.1132\times10^{-9}\). Red curves represent non-zero axion DM fraction $ f_\textrm{ax}$ with $m_\textrm{ax} = 10^{-24}$ eV. Turquoise curves represent reduced $A_\mathrm{s}$ relative to the fiducial cosmology, mimicking the suppression of structure caused by axion DM. Both effects, by suppressing the initial matter power spectrum, lead to an increase in the number of low-mass halos at late times, due to the delay in the onset of hierarchical growth.}
    \label{fig:HMF_As}
\end{figure}

Figure \ref{fig:HMF_m25_z} shows that the suppression of the HMF due to axion DM is more pronounced at higher redshift, motivating the use of JWST and other high-redshift measurements of the UVLF. Fig.~\ref{fig:HMF_m25_z} shows the ratio of the 10\%-axion HMF to the \(\Lambda\)CDM HMF for an axion mass of $10^{-25}$ eV at the same redshift. For $z=12$, the HMF is suppressed by more than five orders of magnitude for halos of mass $10^{13} M_\odot$. However, for lower redshifts such as $z=0$, the maximum suppression is only around $\sim 25\%$. The halo suppression is stronger at earlier times since the halos are a cleaner probe of the primordial axion suppression. At later times, non-linearities tend to erase the axion suppression. In fact, for $z \leq 3$, we see more low mass halos for 10\% axion models relative to \(\Lambda\)CDM. As mentioned above, this enhancement effect is due to the delay in hierarchical growth of halos, resulting in an overabundance of low-mass halos at late times, as shown in the lower panel of Fig.~\ref{fig:HMF_m25_z}. Fig.~\ref{fig:HMF_As} shows that reducing the primordial scalar amplitude $A_\mathrm{s}$ results in a similar overabundance of low-mass halos at \(z = 0\), which is also due to the suppression of structure leading to a delay in hierarchical growth. Although, the scale dependence of the effect is different in the case of axions due to the additional Jeans suppression. Axion enhancement effects are only relevant for very low halo masses ($M_h \lesssim 10^9 M_\odot$) and late times ($z \lesssim 3$), which are well beyond the scope of the UVLF (see Fig.~\ref{fig:Hubble_range}). This overabundance of low-mass halos in a mixed axion cosmology is nonetheless a novel theoretical result. We discuss the potential observability with low-\(z\) probes in Sec. \ref{sec:Discussion}.

%Figure \ref{fig:HMF_m25_z} also illustrates the effect of axion physics on the HMF below the cutoff mass of $10^{9.5} M_\odot$ (as calculated in \cite{Vogt:2022bwy}). Halos below this cutoff can only form from the CDM distribution, as it is assumed that axions do not cluster below this scale.

\section{The Galaxy UV Luminosity Function}
\label{sec:UVLF}

\begin{figure}
    \centering
    \includegraphics[trim={0cm 1cm 1cm 1cm }, clip,width=\linewidth]{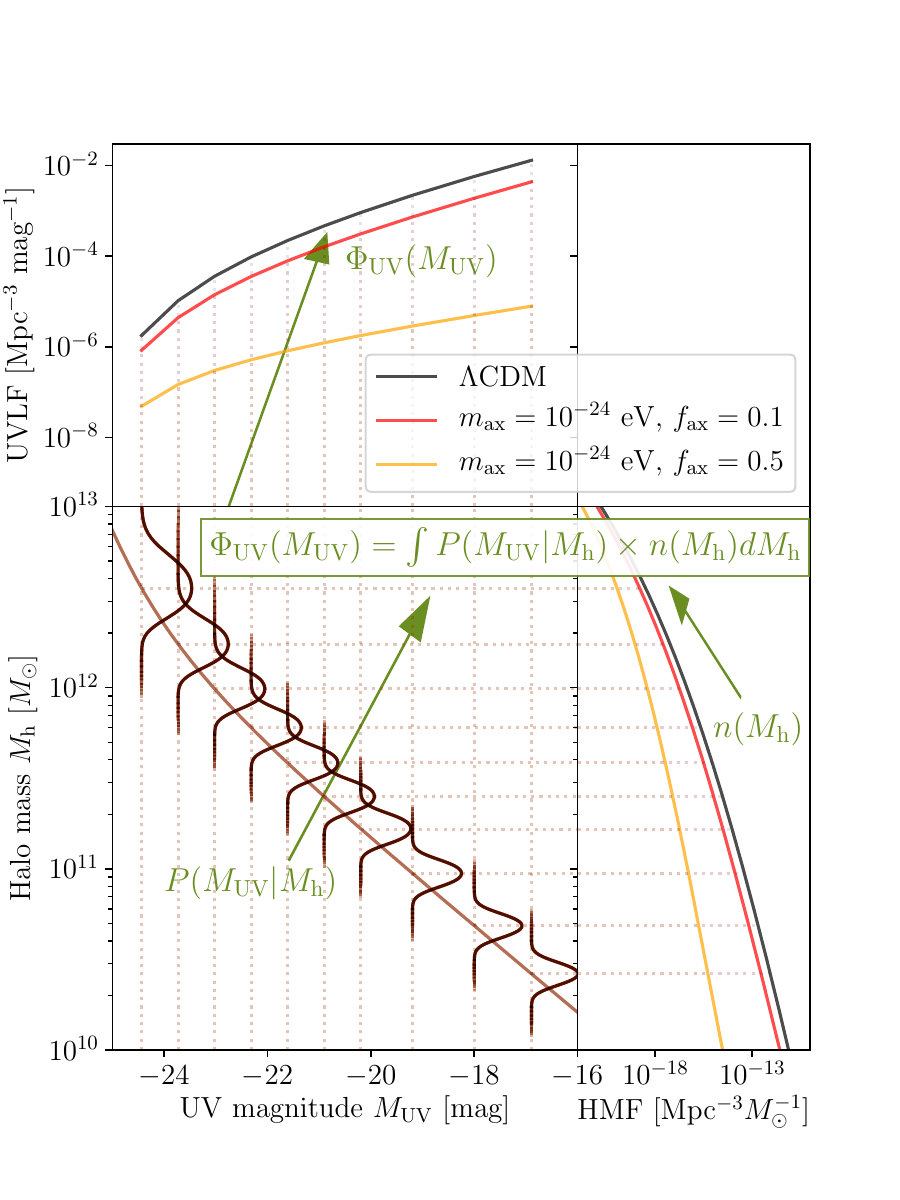}
    \caption{A schematic illustrating how the UVLF ($\Phi_\mathrm{UV}(M_\mathrm{UV})$, upper left) can be computed from the HMF ($n(M_h)$, lower right) and the average UV magnitude $\overline{M_\mathrm{UV}}$ for halos of a given mass $M_h$ (lower left), following Eq.~\eqref{eq:UVLF_construct}. The solid Gaussian-like curves in the lower left panel represent the conditional probability of a halo having mass $M_\mathrm{h}$ given UV magnitude $M_\mathrm{UV}$ (i.e., the inverse of the probability given in Eq.~\eqref{eq:UVprob}, as this better illustrates the construction of the UVLF by integrating over multiple distributions of Eq.~\eqref{eq:UVprob} for different values of \(M_h\)). We show a pure CDM universe (in black), a mixed-DM universe with 10\% axions and $ m_\textrm{ax} = 10^{-24}$ eV in red, and one with 50\% axions of the same mass in orange. All scenarios are plotted for $z=6$. Axions impact the HMF (lower right), and thus the UVLF (upper left), but do not impact the average UV magnitudes for halos of a given mass (lower left).}
    \label{fig:UVLF_HMF_grid}
\end{figure}

\subsection{Computing the UVLF} 
\label{sec:UVLF_compute}

The UVLF $\Phi_\mathrm{UV}$ is the number density of galaxies with a given UV magnitude $M_\mathrm{UV}$ at a redshift $z$. The UVLF is dependant on the number density of galaxy halos with a given mass (the HMF $n(M_h, z)$ described in Section \ref{sec:HMF}) and the average UV magnitude of galaxies in halos with a given mass. This relationship can be expressed as
\begin{equation}
    \Phi_\mathrm{UV}(M_\mathrm{UV}, z) = \int n(M_h,z) P(M_\mathrm{UV} | M_h, z ) dM_h,
    \label{eq:UVLF_construct}
\end{equation}
where $P(M_\mathrm{UV} | M_h, z )$ is the probability of a galaxy having a UV magnitude $M_\mathrm{UV}$ given a halo mass $M_h$ at redshift $z$ (using the conditional luminosity function formalism developed by \cite{Yang:2002ww} and implemented in \texttt{GALLUMI} by \cite{Sabti:2021xvh}). This probability can be modeled as a Gaussian distribution:
\begin{multline}
\label{eq:UVprob}
    P(M_\mathrm{UV} | M_h, z ) = \frac{1}{\sqrt{2 \pi} \sigma_{M_\mathrm{UV}}} \\ 
    \times \exp{\bigg[ - \frac{\big(M_\mathrm{UV} - \overline{M_\mathrm{UV}}(M_h, z)\big)^2}{\sigma^2_{M_\mathrm{UV}}}\bigg]},
\end{multline}
where $\sigma_{M_\mathrm{UV}}$ is the scatter in UV magnitude (driven by, e.g., bursty star formation), which we treat as a nuisance parameter that we fit to the data. $\overline{M_\mathrm{UV}}(M_h, z)$ is the average UV magnitude of galaxies in halos with mass $M_h$ at redshift $z$, which is computed according to the model that we now specify. The relation given in Eq.~\eqref{eq:UVprob} is illustrated in Fig.~\ref{fig:UVLF_HMF_grid}.
%This average UV magnitude is identically related to the UV luminosity $L_\mathrm{UV}$ by the formula,
%\begin{equation}
%    \log_{10}\bigg(\frac{L_\mathrm{UV}}{\text{erg s}^{-1}}\bigg) = 0.4 \big( 56.63 - M_\mathrm{UV} \big).
%\end{equation}

We relate the mean UV magnitude $\overline{M_\mathrm{UV}}$ to the halo mass $M_h$ using models presented in \cite{Sabti:2021xvh}.\footnote{The method described here is denoted as model II in \cite{Sabti:2021xvh}. Appendix \ref{sec:AppB} contains a discussion and comparison to their model III, while model I is not used as it depends on a description of halo accretion that is impacted in non-trivial ways by axion physics.} First, the mean UV luminosity $\overline{L_\mathrm{UV}}$ is related to the mean UV magnitude by the definition
\begin{equation}
\label{eq:luminosity}
    0.4(51.63 - \overline{M_\mathrm{UV}}) = \log_{10}\bigg( \frac{\overline{L_\mathrm{UV}}}{\text{erg s}^{-1}} \bigg).
\end{equation}

The mean UV luminosity $\overline{L_\mathrm{UV}}$ can be related to the mean star formation rate $\overline{\dot{M}_\star}$:
\begin{equation}
    \overline{L_\mathrm{UV}} = \frac{\overline{\dot{M}_\star}}{ \kappa_\mathrm{UV}} ,
    \label{eq:UV_lum}
\end{equation}
where $\kappa_\mathrm{UV} = 1.15 \times 10^{-28} M_\odot$ s erg$^{-1}$ yr$^{-1}$ is derived from stellar synthesis population modeling (see \cite{Madau:2014bja} for details). This relation physically captures how UV photons are generated in young star-forming regions. 

We can then relate the mean star formation rate to the mean stellar mass $\overline{M_\star}$:
\begin{equation}
    \label{eq:dynamic}
    \overline{\dot{M}_\star} = \frac{H(z)}{t_\star} \overline{M_\star},
\end{equation}
which assumes the stellar accretion rate is proportional to the dynamical time of DM halos, as done in \cite{Park:2018ljd} and \cite{Gillet:2019fjd}. This relationship is parameterized by the dimensionless $t_\star$ which we vary as a free parameter. \rev{In Appendix~\ref{sec:AppMS}, we compare this linear relation to measurements of the galaxy main sequence \citep{Popesso2023}.}

Finally, we relate the average stellar mass $\overline{M_\star}$ to the halo mass $M_h$ using a parametric broken power law equation:
\begin{equation}
    \frac{\overline{M_\star}}{M_h} = \frac{\epsilon_\star}{\left(\frac{M_h}{M_c}\right)^{\alpha_\star} + \left(\frac{M_h}{M_c}\right)^{\beta_\star}},
    \label{eq:double_power}
\end{equation}
where $\epsilon_\star$, $M_c$, $\alpha_\star$, and $\beta_\star$ are functions over which we marginalize when fitting to data. $\epsilon_\star \geq 0$ is the overall amplitude of stellar mass, $M_c \geq 0$ is the mass at which mean stellar mass peaks relative to halo mass, $\alpha_\star \leq 0$ regulates the slope of the low-mass end of the function, while $\beta_\star \geq 0$ regulates the high-mass end. This broken power law accurately reflects the physical model of star formation, since the galaxies in low-mass halos will have insufficient star formation, while those in high mass halos will have used up most of the available gas. 
\rev{This relation agrees with current measurements of the stellar mass function, such as \cite{2024MNRAS.tmp.1880W}}. 
The exact slope of the two effects described above and the pivot mass at which star formation peaks, are all free parameters in this model. Following the parameterization in \cite{Sabti:2021xvh}, we assume the $z$-dependence of these parameters is:
\begin{align}
\label{eq:variables}
    \alpha_\star(z) &= \alpha_\star, \\
    \beta_\star(z) &= \beta_\star, \\
    \log_{10}\epsilon_\star(z) &= \epsilon_\star^s \times \log_{10}\bigg(\frac{1+z}{1+6}\bigg) + \epsilon_\star^i, \\
    \log_{10} M_c(z) &= M_c^s \times \log_{10}\bigg(\frac{1+z}{1+6}\bigg) + M_c^i,
    \label{eq:variables_end}
\end{align}
where both $\epsilon_\star$ and $M_c$ are parameterized by a power law with slopes $[\epsilon_\star^s, M_c^s]$ and intercepts $[\epsilon_\star^i, M_c^i]$, both pivoting at $z=6$ (chosen by \cite{Sabti:2021unj} to be representative of the HST UVLF estimates, shown in Fig.~\ref{fig:Hubble_range}). Since $\epsilon_\star$ is fully degenerate with $t_\star$ from Eq.~\eqref{eq:dynamic}, we combine $t_\star$ into our nuisance function $\epsilon_\star (z)$ (as done in \cite{Sabti:2021xvh}).

\begin{figure}
    \centering
    \includegraphics[trim={0.3cm 0.2cm 1cm 0.2cm }, clip,width=\linewidth]{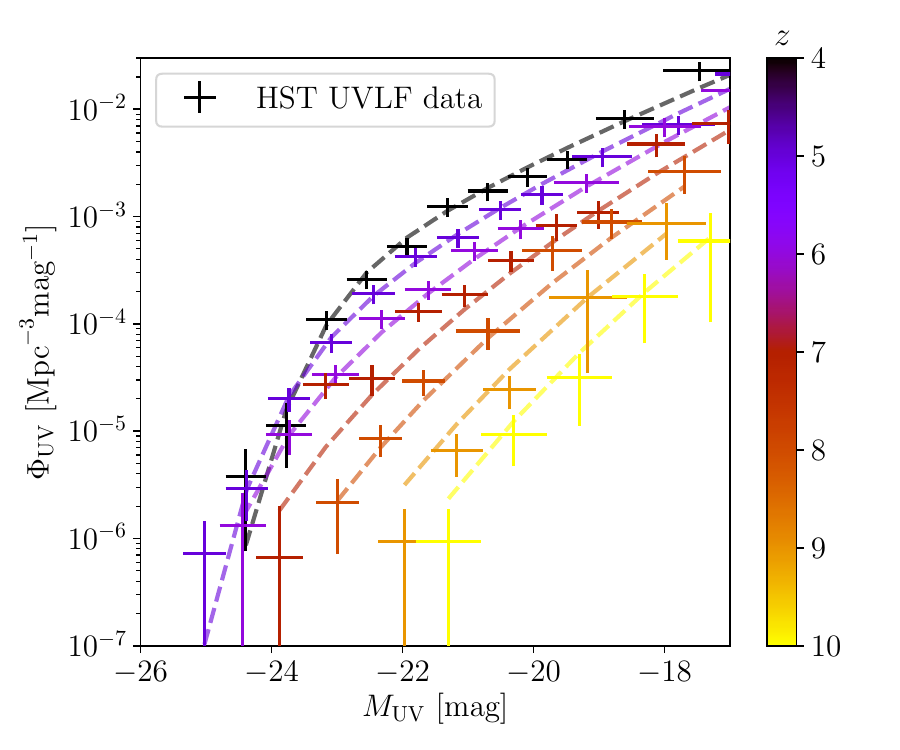}
    \caption{The UVLF for redshifts $z=4-10$, assuming a pure CDM cosmology fit to Hubble Space Telescope (HST) data, alongside the HST data showing 68\% c.l. uncertainties and magnitude bin widths.}
    \label{fig:UVLF_fiducial_z}
\end{figure}

When combined together, Eqs.~\eqref{eq:luminosity} to \eqref{eq:variables_end} allow us to compute the average UV luminosity for a given halo mass, which we then use to calculate the UVLF [Eqs.~\eqref{eq:UVLF_construct} and \eqref{eq:UVprob}]. \cite{Sabti:2021xvh} demonstrates that the model we present above can reproduce the UVLF as calculated in the detailed Illustris TNG hydrodynamical simulations \citep{IllustrisTNG_JWST}. This test means that we can marginalize over uncertainty in the relation between UV luminosity and halo mass as seen in hydrodynamical simulations, but without the need to run a large number of computationally-expensive models. We discuss below the consequences of axions on this astrophysical model. The results of this model (with HST best-fit parameters) are shown in Fig.~\ref{fig:UVLF_fiducial_z} along with HST data.
%When combined together, this gives us a model for computing the UVLF, by first computing the average UV magnitude for halos of a given mass, and then integrating over the probability of observing a certain UV luminosity from halos of any mass, multiplied by the number density of halos with that mass (the HMF). This gives us an estimate of the expected number of UV sources of a given magnitude at a given redshift.

%bar all the mstar

%F \geq o0l ilowing the method in [CITE GALLUMI PAPER], we relate the UV luminosity $L_\mathrm{UV}$ to the star formation rate $\dot{M}_\star$ by the formula
%\begin{equation}
%    L_\mathrm{UV} = \frac{\dot{M}_\star}{\kappa_\mathrm{UV}} ,
%\end{equation}
%where the conversion factor $\kappa_\mathrm{UV} = 1.15 \times 10^{-28} M_\odot \text{s erg}^{-1} \text{yr}^{-1}$ is derived from a stellar synthesis population model [CITE THE SAME PAPER CITED IN GALLUMI].

%\subsection{Impact of axions on the UVLF}

\begin{figure}
    \centering
    \includegraphics[trim={0.1cm 1.8cm 1.4cm 2.2cm }, clip,width=\linewidth]{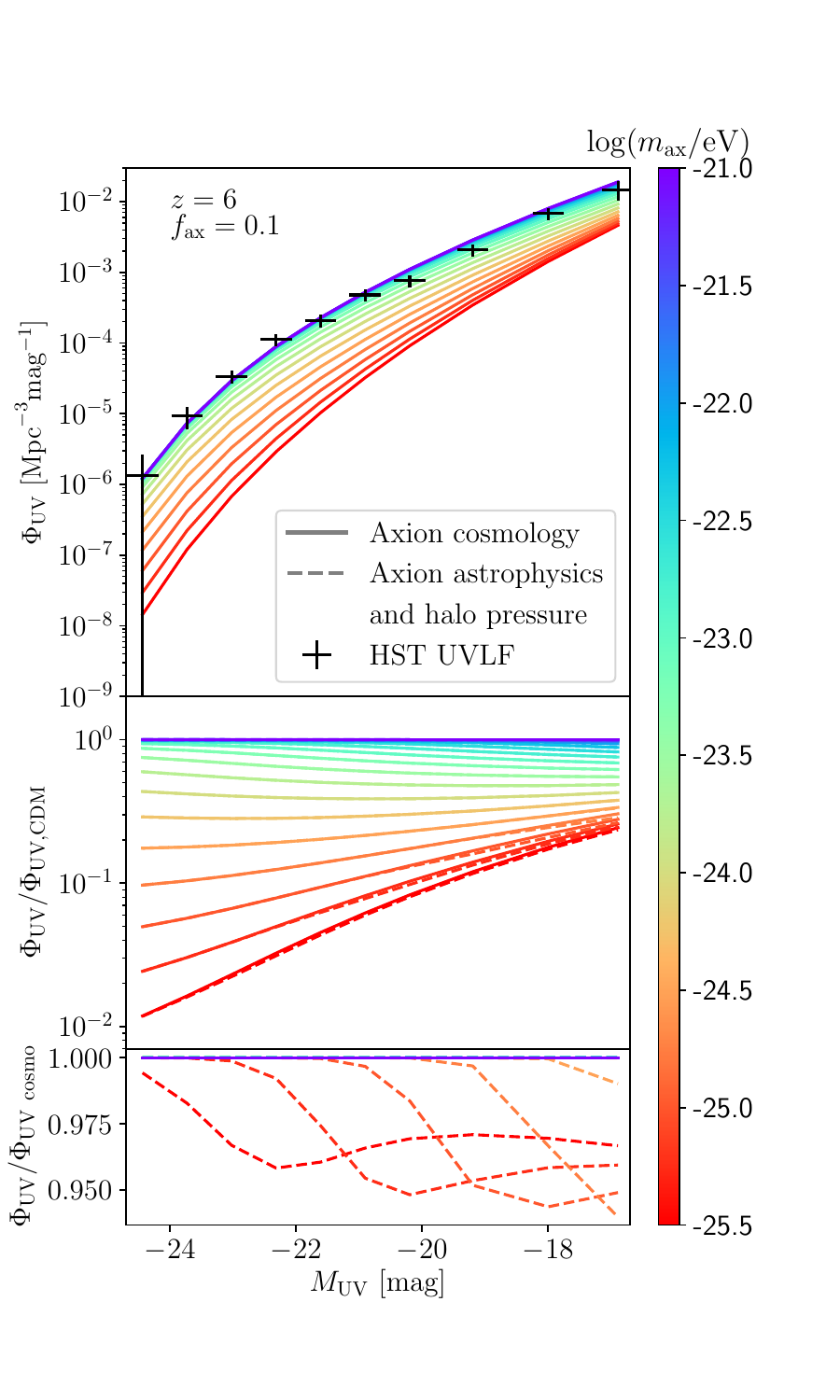}
    \caption{The UVLF for a range of axion masses, all with a 10\% axion fraction and at $z=6$. Black error bars show HST data at the same redshift (68\% c.l. and magnitude bin widths). The upper panel shows the cosmological impact of axion mass on the UVLF without considering axion astrophysics. The middle panel shows the impact of both axion cosmology (in solid lines) and axion halo pressure and astrophysics (in dashed lines) relative to the pure CDM model. The UVLF is more suppressed for axions with lower mass, due to the larger de Broglie wavelength. Higher mass axions cause more suppression at lower halo masses (and hence fainter galaxies), while lower mass axions cause more suppression at higher halo masses (and hence brighter galaxies). This effect is due to the same overabundance of low-mass halos shown in Figs.~\ref{fig:HMF_axions}, \ref{fig:HMF_m25_z} and \ref{fig:HMF_As}. The lowest panel shows the impact of both axion halo pressure and axion astrophysics relative to the UVLF with only the cosmological impacts. Axion halo pressure and astrophysics only impact the UVLF for $m_\textrm{ax} \lesssim 10^{-24.5}\mathrm{eV}$, with about 5\% additional suppression, whereas axion cosmology suppresses the UVLF by one to two orders of magnitude for similarly low axion masses.}
    \label{fig:UVLF_mass}
\end{figure}

In addition to the effects of axions on the halo mass function discussed in Section \ref{sec:HMF}, we consider here the effects of axion astrophysics on the relation between UV luminosity and halo mass. As in the discussion regarding axion halo pressure, axions do not cluster into halos below a critical mass equivalent to the Jeans scale within the halo \citep{Vogt:2022bwy}. Therefore, the relation between halo mass and stellar mass (Eq. \eqref{eq:double_power}) can deviate from the CDM expectation below this critical mass. Halos in that regime are missing the axion DM component and thus have a higher ratio of baryons to DM. For a given stellar mass, the halo mass is multiplied by a factor equal to the CDM fraction, i.e. by $(1 - \Omega_\mathrm{ax}/\Omega_\mathrm{DM})$, where $\Omega_\mathrm{ax}$ and $\Omega_\mathrm{DM}$ are the axion and total DM densities, respectively, relative to the critical density. This reduction in halo mass means that the ratio of the stellar mass relative to the total halo mass, $\overline{M_\star} / \overline{M_h}$, is enhanced by a factor of $1/(1 - \Omega_\mathrm{ax}/\Omega_\mathrm{DM})$. This only applies for halos below the cutoff mass given in Eq.~\eqref{eq:jeans_mass}, modifying Eq.~\eqref{eq:double_power} with a piecewise enhancement below the critical mass. We illustrate the negligible role of this ``axion astrophysics'' in Fig.\ref{fig:UVLF_mass}. We discuss the UVLF model and how it may change in the presence of axions further in Section \ref{sec:Discussion}.
%while including them with approximate models could produce constraints that are deceptively tight, depending on the accuracy of those models. Therefore, in this work we neglect the effects of axion astrophysics on the UVLF, since the effects are subdominant and difficult to model, and neglecting them results in slightly conservative constraints.

%we conclude that it is safe to ignore axion astrophysical effects of axions when studying the UVLF in this regime of luminosities (although we do note that this might change if we were able to extend the UVLF to fainter luminosities, corresponding to lighter halos).

\subsection{Impact of axions on the UVLF}
\label{sec:axions_UVLF}

 \begin{figure*}
    \centering
    \includegraphics[trim={2.2cm 1.5cm 3cm 2cm }, clip,width=\linewidth]{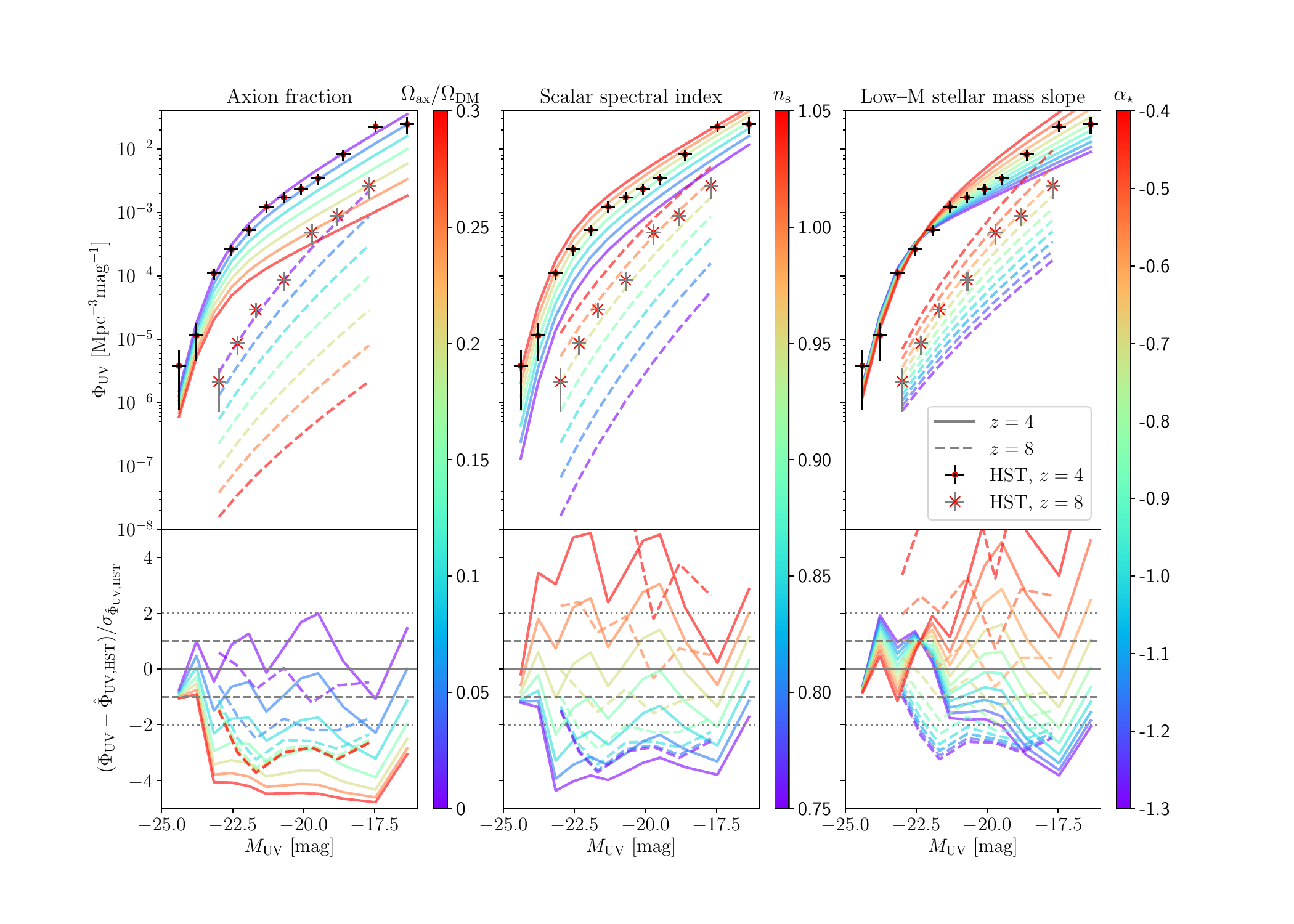}
    \caption{The impact on the UVLF (all other parameters fixed at best-fit values for HST UVLF + \textit{Planck} CMB) from axion fraction (for $ m_\textrm{ax} = 10^{-24}\,\mathrm{eV}$), primordial scalar spectral index $n_\mathrm{s}$ and the low-mass slope of the stellar mass - halo mass relation $\alpha_\star$. The UVLF is shown at $z=4$ and $z=8$, alongside the HST measurements at the same redshifts (respectively black and grey error bars; 68\% c.l. uncertainties and magnitude bin widths). The lower panels show the model difference to the HST data rescaled by the data uncertainties. All three variables have qualitatively similar impacts on the UVLF, meaning that a higher axion fraction is degenerate with higher $n_s$ and/or $\alpha_\star$.}
    \label{fig:UVLF_HMF_3var}
    %cut, and put first column added to fig 11, to show both mass and fraction
\end{figure*}

Fig.~\ref{fig:UVLF_mass} shows that axions suppress the UVLF due to the suppression of structure on small scales as captured in the HMF. Lower mass axions tend to suppress structure more. Higher mass axions cause more suppression at faint luminosities, while low mass axions (\(m_\mathrm{ax} < 10^{-22.5}\,\mathrm{eV}\)) cause more suppression at higher luminosities. This is due to the same overabundance of low-mass halos shown in Figures \ref{fig:HMF_axions}, \ref{fig:HMF_m25_z}, and \ref{fig:HMF_As}.
%but that the magnitude-dependence of the suppression is not as pronounced as the mass-dependence of the impacts on the HMF. This is likely due to the spread in UV magnitudes, $\sigma_{M_\mathrm{UV}}$, resulting in the HMF at a certain halo mass impacting the UVLF at many UV magnitudes. This is illustrated in Fig.~\ref{fig:UVLF_HMF_grid}.

The first column in Fig.~\ref{fig:UVLF_HMF_3var} shows how an axion with $m_\textrm{ax} = 10^{-24}$ eV impacts the UVLF at different axion fractions. Increasing the axion density increases the suppression of faint UV galaxies, which are typically hosted by lower mass halos. However, the impact is qualitatively similar to varying the primordial scalar spectral index $n_\mathrm{s}$ (shown in column 2 of Fig.~\ref{fig:UVLF_HMF_3var}) or by varying the the low-mass slope of the stellar mass - halo mass relation $\alpha_\star$ (in the third column). These degeneracies mean that higher axion fractions can be permitted by increasing $n_\mathrm{s}$ and/or $\alpha_\star$, i.e., by allowing for more primordial small-scale structure or by increasing the stellar to halo mass ratio at the low mass end. Further, the top panels show that the axion effect is relatively stronger at higher redshift, driven by the effect on the HMF shown in Fig.~\ref{fig:HMF_m25_z}. However, the lower panels show that HST data, which have fewer high-\(z\) sources, are less constraining at higher redshift. We discuss in Section \ref{sec:Discussion} the utility of higher redshift observations from JWST.
%\clearpage

% making sure there isn't a hanging section title
%\vspace{0.3in}
\section{Comparison to Observations}
\label{sec:MCMC}

\begin{table}[ht]
 \begin{center}
   \begin{tabular}{c  c  c }
    \hline
    \hline
    Shorthand & Likelihoods and priors & Color \\
    \hline
    HST UVLF & HST UVLF + Pantheon  & \textcolor{HWblue}{Blue} \\
    & + BBN $\omega_\mathrm{b}$ prior  & \\
    &  + fixed $\theta_\mathrm{s}$ + fixed $\tau$ & \\
    \hline
    \textit{Planck} CMB & \textit{Planck} high-$\ell$ TT, TE, EE & \textcolor{HWgreen}{Green}\\
    & +  \textit{Planck} low-$\ell$ TT, EE & \\
    & + \textit{Planck} lensing + BOSS BAO & \\
    \hline
    HST UVLF &  HST UVLF + Pantheon & \textcolor{HWred}{Red} \\
    + \textit{Planck} CMB & + \textit{Planck} high-$\ell$ TT, TE, EE & \\
    &  + BBN $\omega_\mathrm{b}$ prior &\\
    & +  \textit{Planck} low-$\ell$ TT, EE & \\
    & + \textit{Planck} lensing + BOSS BAO & \\
    \hline
    JWST UVLF & JWST UVLF + Pantheon  & \textcolor{HWpink}{Pink} \\
    & + BBN $\omega_\mathrm{b}$ prior & \\
    &+ fixed $\theta_\mathrm{s}$ + fixed $\tau$ & \\
    \hline
    HST  & HST UVLF + JWST UVLF   & \textcolor{HWmagenta}{Magenta} \\
    + JWST joint & + Pantheon + BBN $\omega_\mathrm{b}$ prior & \\
    & + fixed $\theta_\mathrm{s}$ + fixed $\tau$ & \\
    \hline
    \end{tabular}
    \caption{The different combinations of likelihoods, priors and fixed parameters that we use, along with the shorthand used in the text and the color used for contour plots. HST UVLF measurements are given in \cite{HST_UVLF}, Pantheon supernovae magnitudes are given in \cite{Pan-STARRS1:2017jku}, all \textit{Planck} CMB likelihoods are given in \cite{Planck:2018vyg}, BOSS BAO measurements are given in \cite{BOSS:2013rlg} and JWST UVLF measurements are given in \cite{JWST_UVLF}. The prior on $\omega_\mathrm{b}$ and the fixed $\theta_\mathrm{s}$ and $\tau$ values are described in Sec.~\ref{sec:HST_UVLF}.}
    \label{Tab:constraint_combos}
    %make it clear that this is just for 10^-24, reference figure 18 for mass dependence
    \end{center}
\end{table}   

We now use UVLF data from HST, combined with other cosmological observations, to compute constraints on mixed axion models. We use the Markov chain Monte Carlo (MCMC) code \texttt{MontePython} \citep{Brinckmann:2018cvx}, along with the \texttt{axiCLASS} axion Boltzmann solver \citep{Poulin:2018dzj, Smith:2019ihp} based on the \texttt{CLASS} code \citep{CLASS}, to sample the posterior distribution. We use a modified version of the \texttt{GALLUMI} UVLF likelihood (presented in \cite{Sabti:2021xvh}) to compare our axion models to HST measurements of the UVLF (see Section \ref{sec:HST_UVLF}). We also use \textit{Planck} measurements of the CMB in order to constrain larger scales in the matter power spectrum than we probe in the UVLF. We summarize the different data combinations we consider in Table \ref{Tab:constraint_combos}, including the JWST data we consider in Section \ref{sec:Discussion}. All posterior corner plots are generated using the \texttt{corner} package \citep{corner}.

\subsection{Hubble UV luminosities}
\label{sec:HST_UVLF}

We use estimates of the UVLF from the Hubble Space Telescope (HST) from \cite{HST_UVLF}. This sample contains $>24,000$ UV sources, at redshifts $z=4$ to $z=10$. The UV sources are grouped into $M_\mathrm{UV}$ bins of 0.5 magnitudes. These estimates of the UVLF are shown in Fig.~\ref{fig:UVLF_fiducial_z}.

We use the \texttt{GALLUMI} likelihood \citep[presented in][]{Sabti:2021xvh} to compute the UVLF and compare it to the HST estimates using a Gaussian likelihood $\mathcal{L}$:
\begin{equation}
\log(\mathcal{L}) = \sum_{M_\mathrm{UV},z} \bigg(\frac{\Phi_\mathrm{UV}(M_\mathrm{UV},z) - \hat{\Phi}_\mathrm{UV,HST}(M_\mathrm{UV},z)}{\sigma_{\Phi_\mathrm{UV,HST}}(M_\mathrm{UV},z)}\bigg)^2,
\end{equation}
where $\Phi_\mathrm{UV}(M_\mathrm{UV},z)$ is the UVLF computed using Eq. \eqref{eq:UVLF_construct} and $\hat{\Phi}_\mathrm{UV,HST}(M_\mathrm{UV},z)$ and $\sigma_{\Phi_\mathrm{UV,HST}}(M_\mathrm{UV},z)$ are the HST measurements of the value and uncertainty, respectively, of the UVLF at redshift $z$ and UV magnitude $M_\mathrm{UV}$. We vary all seven UVLF model parameters $[\alpha_\star, \beta_\star, \epsilon_\star^s, \epsilon_\star^i, M_c^s, M_c^i, \sigma_{M_\mathrm{UV}}]$ (defined in Eqs.~\eqref{eq:UVprob} and \eqref{eq:double_power} - \eqref{eq:variables_end}) as astrophysical nuisance parameters with uniform priors (see Table \ref{Tab:param_priors}).

Following \cite{Sabti:2021xvh}, we combine the HST UVLF likelihood with supernovae magnitudes from the Pantheon survey 
%in order to provide a measurement of the total matter density $\Omega_\mathrm{m}$
\citep[using the Pantheon likelihood presented in][]{Pan-STARRS1:2017jku}. However, we do not calibrate the supernovae magnitudes with Cepheid variable star data. This likelihood adds one additional nuisance parameter $M$, a calibration magnitude related to the intrinsic supernova luminosity, which we allow to vary as a free parameter with a uniform prior (see Table \ref{Tab:param_priors}). We also impose a Gaussian prior on $\omega_b\times10^2 = \Omega_b h^2\times10^2$ of $\mathcal{N}(2.233, 0.036)$ from Big Bang nucleosynthesis (BBN) measurements \citep{Pisanti:2020efz} and fix the values of the acoustic angular scale $100 \theta_s = 1.040827$ and optical depth of reionization $\tau = 0.0544$ to be consistent with \textit{Planck} values \citep{Planck:2018vyg}. Including both the Pantheon supernovae data with the \textit{Planck} value of $100 \theta_s = 1.040827$ tightly constrains both the total matter density $\Omega_\mathrm{m}$ and the dimensionless Hubble parameter $h$.

\begin{figure}
    \noindent
    \includegraphics[trim={0.1cm 0.1cm 0.5cm 0.5cm }, clip,width=1.0\linewidth]{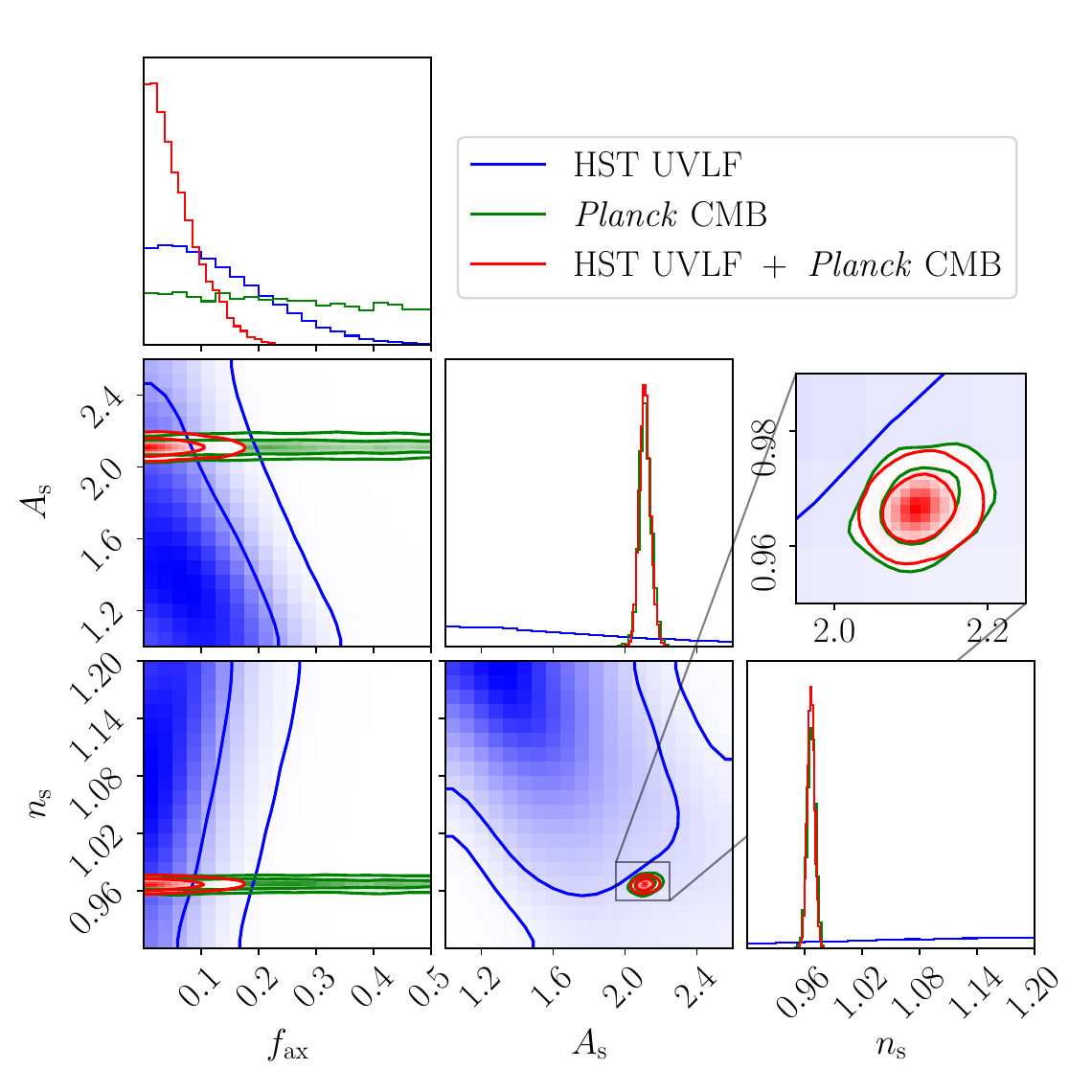}
    \caption{Marginalized posterior of axion fraction $ f_\textrm{ax} = \Omega_\mathrm{ax}/\Omega_\mathrm{DM}$ for axion mass of $10^{-24}$ eV, the primordial scalar amplitude $A_\mathrm{s}$ and spectral index $n_\mathrm{s}$. The \textcolor{HWblue}{blue} contours are from HST UVLF data shown in Fig.~\ref{fig:UVLF_fiducial_z} (plus Pantheon supernovae data and a BBN prior on $\omega_\mathrm{b}$, as explained in Table~\ref{Tab:constraint_combos}). The \textcolor{HWgreen}{green} contours are from \textit{Planck} CMB plus BAO data from BOSS. The \textcolor{HWred}{red} contours are the joint constraints from both sets of observables. \(A_\mathrm{s}\) is given in units of \(10^{-9}\).}
    \label{fig:corner_joint_cosmo}
\end{figure}

%\subsection{James Webb UV Luminosities}

%\subsection{Results}
%#\label{sec:Results}

\subsection{\textit{Planck} cosmic microwave background}

We use \textit{Planck} measurements of the cosmic microwave background (CMB) in order to anchor the values of cosmological parameters, in particular, $A_s$ and $n_s$ as these are measures of the large-scale matter power that the UVLF does not probe. We use the \textit{Planck} 2018 foreground-marginalized likelihood, using the high-multipole $\ell$ TT, TE and EE angular power spectra, along with the low-$\ell$ TT and EE angular power spectra \citep{Planck:2019nip}. We also use CMB lensing \(\phi \phi\) power spectra and measurements of the galaxy baryon acoustic oscillations (BAO) from the Baryon Oscillation Spectroscopic Survey (BOSS 2014) \citep{BOSS:2013rlg}. For constraints using \textit{Planck} CMB data, we allow $100 \theta_s$ and $\tau$ to vary as cosmological parameters and also vary the calibration parameter $A_{\textrm{Planck}}$ \citep{Planck:2018vyg}. We use uniform priors given in Table \ref{Tab:param_priors}.

\subsection{Results}
\label{sec:Results}

In Fig.~\ref{fig:corner_joint_cosmo}, we show the marginalized posterior of the axion fraction $ f_\textrm{ax} = \Omega_\mathrm{ax}/\Omega_\mathrm{DM}$, the primordial scalar amplitude $A_\mathrm{s}$ and spectral index $n_\mathrm{s}$, for an axion of mass $10^{-24}$ eV. We illustrate the degeneracy between axion and cosmological parameters at this fixed mass as the power of adding UVLF data is clear. The \textit{Planck} likelihood (combined with BOSS BAO, as described in Tab.~\ref{Tab:constraint_combos}) limits the axion fraction 
\begin{eqnarray}
\label{eq:result_Planck}
f_\mathrm{ax} [m_\mathrm{ax}=10^{-24} \mathrm{eV}] &<& 0.93 \,\,\mathrm{(95\% \ C.L.)} \nonumber \\ &&[Planck\mathrm{\ CMB}]. \nonumber\\
\end{eqnarray} 
This limit is consistent with the \textit{Planck} + BOSS galaxy clustering constraints presented in \cite{Rogers:2023ezo}. The HST UVLF likelihood already sets a stronger constraint, limiting the axion fraction 
\begin{eqnarray}
\label{eq:result_HST}
f_\mathrm{ax} [m_\mathrm{ax}=10^{-24} \mathrm{eV}] &<& 0.28 \,\,\mathrm{(95\% \ C.L.)} \nonumber \\ &&[\mathrm{HST}~\mathrm{UVLF}]. \nonumber\\
\end{eqnarray} 
The UVLF is a more powerful probe, since for \(m_\mathrm{ax} \geq 10^{-25}\,\mathrm{eV}\), the axion wavelength is smaller than the smallest modes currently modeled in \textit{Planck} (and other CMB experiments') data. Whereas, as illustrated in Fig.~\ref{fig:power_spectrum}, the UVLF probes much smaller scales than current CMB data where axion effects manifest.

The lower-left panel of Fig.~\ref{fig:corner_joint_cosmo} shows that for the UVLF constraints, slightly higher values of $ f_\textrm{ax}$ are allowed with higher values of $n_\mathrm{s}$\footnote{Larger axion fractions can be consistent with the HST UVLF when combined with a higher $n_\mathrm{s}$, as $n_\mathrm{s}$ raises small-scale power while the axion fraction suppresses it. The impact of $\Omega_\mathrm{ax}$ and $n_\mathrm{s}$ can be compared in Fig.~\ref{fig:UVLF_HMF_3var}.}. High values of $n_\mathrm{s}$ are strongly inconsistent with \textit{Planck} CMB data, which constrain $n_\mathrm{s} = 0.9665 \pm 0.0038$ \citep{Planck:2019nip}. Therefore, if we combine the HST UVLF measurements of smaller-scale structure with the \textit{Planck} CMB large-scale measurements of $A_\mathrm{s}$ and $n_\mathrm{s}$, we break the degeneracy between axion and cosmological parameters. We thus significantly improve our axion fraction limit 
\begin{eqnarray}
f_\mathrm{ax} [m_\mathrm{ax}=10^{-24} \mathrm{eV}] &<& 0.15 \,\,\mathrm{(95\% \ C.L.)} \nonumber \\ &&[\mathrm{HST}~\mathrm{UVLF}+Planck\mathrm{\ CMB}]. \nonumber\\
\end{eqnarray} 
%at 95\% credibility for $m_\mathrm{ax} = 10^{-24}$ eV.

\begin{table}[ht]
 \begin{center}
   \begin{tabular}{ c  c  c }
    \hline
    \hline
    Parameter & Prior & Posterior ($m_\mathrm{ax} = 10^{-24}$ eV) \\
    & & HST UVLF + \textit{Planck} CMB \\
    \hline
    $\Omega_\textrm{CDM}$ & $\mathcal{U}$[0.0, 0.3] & $0.247^{+0.015}_{-0.008}$ \\
    $\Omega_\textrm{ax}$ & $\mathcal{U}$[0.0, 0.3] & $< 0.036$ (95\% c.l.)\\
    $\omega_\mathrm{b}\times 10^2$ & $\mathcal{N}(2.233, 0.036)$ & $2.241^{+0.013}_{-0.013}$\\
    $A_\mathrm{s}\times 10^{9}$ & $\mathcal{U}$[0.5, 3.0] & $2.112^{+0.029}_{-0.033}$\\
    $n_\mathrm{s}$ & $\mathcal{U}$[0.7, 2.0] & $0.9666^{+0.0036}_{-0.0039}$\\
    $100 \times \theta_\mathrm{s}$ & $\mathcal{U}$[1.037, 1.043] & $1.0419^{+0.00028}_{-0.00029}$\\
    $\tau$ & $\mathcal{U}$[0.01, 0.2] & $0.0580^{+0.0069}_{-0.0080}$\\
    %$\log(m_\textrm{ax})$ & $\mathcal{U}$[-26,-21] \\
    \hline
    $\alpha_\star$ & $\mathcal{U}$[-3.0,0.0] & $-0.575^{+0.089}_{-0.114}$\\
    $\beta_\star$ & $\mathcal{U}$[0.0,3.0] & $0.9804^{+0.0024}_{-0.6897}$\\
    $\epsilon_\star^s$ & $\mathcal{U}$[-3.0,3.0] & $0.56^{+0.57}_{-0.33}$\\
    $\epsilon_\star^i$ & $\mathcal{U}$[-3.0,3.0] & $-0.49^{+0.14}_{-0.11}$\\
    $M_c^s$ & $\mathcal{U}$[-3.0, 3.0] & $1.66^{+1.28}_{-0.41}$\\
    $M_c^i$ & $\mathcal{U}$[7.0,15.0] & $11.87^{+0.21}_{-0.15}$\\
    $\sigma_{M_\mathrm{UV}}$ & $\mathcal{U}$[0.001, 3.0] & $0.45^{+0.18}_{-0.19}$\\
    \hline
    $M$ & $\mathcal{U}[-\infty, \infty]$ & $-19.417^{+0.012}_{-0.012}$ \\
    \hline
    $A_\mathrm{Planck}$ & $\mathcal{U}$[0.9,1.1] & $1.0018^{+0.0025}_{-0.0025}$\\
    \hline
    \end{tabular}
    \caption{The priors and posteriors of all cosmological and astrophysical parameters in our joint HST UVLF and \textit{Planck} CMB constraints on mixed axion models with $m_\mathrm{ax} = 10^{-24}$ eV. $\mathcal{U}$ represents uniform priors over the specified range, while $\mathcal{N}$ represents a Gaussian prior with the stated mean and standard deviation. In the right-most column, we give the posterior mean with the 68\% c.l. uncertainties, except where the 95\% c.l. upper limit is given.}
    \label{Tab:param_priors}
    %make it clear that this is just for 10^-24, reference figure 18 for mass dependence
    \end{center}
\end{table}

\begin{figure}
    \noindent
    \includegraphics[trim={0.3cm 0.3cm 1.1cm 1.2cm }, clip,width=\linewidth]{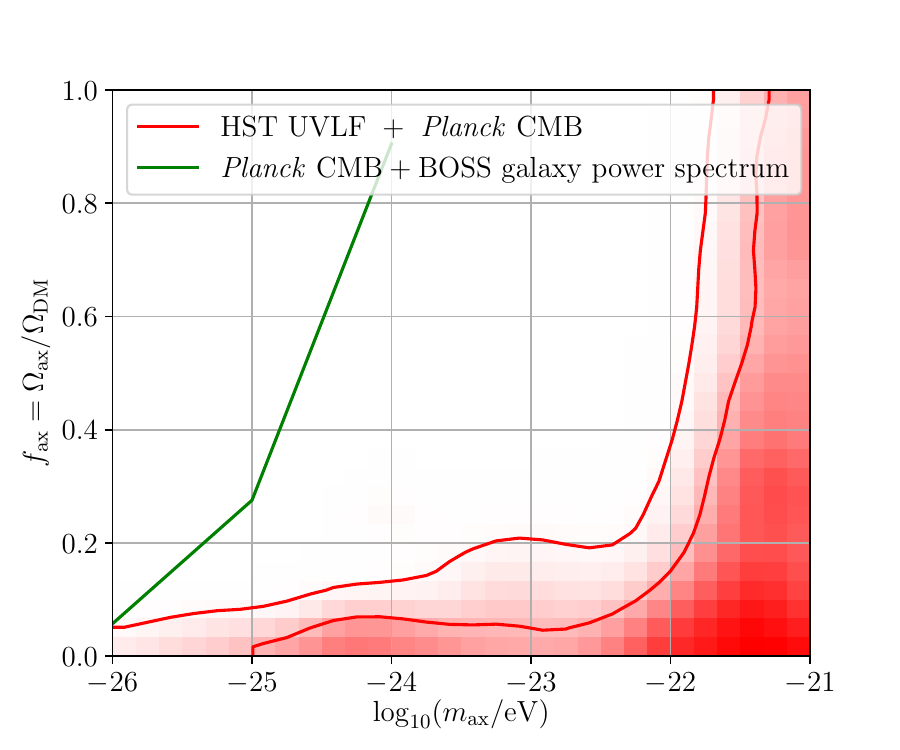}
    \caption{Marginalized posterior on axion mass and DM fraction from joint HST UVLF and \textit{Planck} CMB likelihoods. The \textcolor{HWred}{red} contours indicate the allowed 68\% and 95\% limits, in contrast to the UVLF results in Fig.~\ref{fig:joint_massfrac_compare}. We also show (in \textcolor{HWgreen}{green}) the 95\% upper limit from \textit{Planck} CMB combined with BOSS galaxy power spectrum, as reported in \cite{Rogers:2023ezo}. Combining \textit{Planck} CMB with HST UVLF substantially tightens constraints on the axion fraction.}
    \label{fig:joint_massfrac}
\end{figure}

The constraints shown in Fig.~\ref{fig:corner_joint_cosmo} have a fixed axion mass $ m_\textrm{ax} = 10^{-24}$ eV. However, we also explore how joint HST UVLF and \textit{Planck} CMB likelihoods constrain axions of different masses. In Fig.~\ref{fig:joint_massfrac}, we show constraints on axion fraction given both likelihoods while varying the axion mass $\log(m_\textrm{ax}/\textrm{eV})$ between $-21$ and $-26$, in addition to varying all cosmological and astrophysical parameters as above.\footnote{Following the examples in \cite{Rogers:2020ltq,Dentler:2021zij,Rogers:2023upm}, we set the upper bound on $\log(m_\textrm{ax})$ in a data-informed way, i.e., that the upper bound is set by the sensitivity of the data. In this way, the limits are not affected by the infinite prior volume as $m_\textrm{ax} \rightarrow \infty$.} The 68\% and 95\% upper limits on $f_\mathrm{ax}$ for different axion masses (with a bin width in $\log(m_\mathrm{ax}/\mathrm{eV})$ of $\pm 0.5$) are shown in Table~\ref{Tab:mass_lims}. Axions for $\log(m_\textrm{ax}/\mathrm{eV}) \leq -23$ are limited $\leq 22\%$ of the DM. Axions for $\log(m_\textrm{ax}/\mathrm{eV}) \leq -21.6$ are ruled out as each contributing 100\% of the dark matter at 95\% credibility, thus confirming previous results that exclude axion models as a solution to the so-called cold dark matter ``small-scale crisis'' \citep{Rogers:2020ltq,Nadler:2021dft}. The UVLF loses sensitivity as the axion mass increases, as the axion wavelength falls below the smallest scales probed. We show degeneracies with astrophysical parameters in Appendix \ref{sec:AppA}.

\begin{table}
\centering
\hskip-1.2cm
   \begin{tabular}{ c  c  c }
    \hline
    \hline
     & \multicolumn{2}{c}{Upper limits on $f_\mathrm{ax}$} \\
    %\hline
    $\log(m_\mathrm{ax}/\mathrm{eV})$ & 68\% c.l.  &  95\% c.l. \\
    \hline
-21.0  &  0.620  &  0.930  \\
-22.0  &  0.321  &  0.790  \\
-23.0  &  0.110  &  0.224  \\
-24.0  &  0.071  &  0.151  \\
-25.0  &  0.052  &  0.079  \\
-26.0  &  0.021  &  0.043  \\
     \hline
    \end{tabular}
    \caption{68\% and 95\% c.l. upper limits on the axion DM fraction $f_\mathrm{ax} = \Omega_\mathrm{ax}/(\Omega_\mathrm{ax} + \Omega_\mathrm{CDM})$ for different masses $m_\mathrm{ax}$. These constraints are computed using joint HST UVLF + \textit{Planck} CMB data. The mass-varying posterior chain is divided into bins with width in $\log(m_\mathrm{ax}/\mathrm{eV})$ of $\pm 0.5$. Although the limits at $m_\mathrm{ax} = 10^{-21}$ eV are less than unity, this is expected for a nearly uniform distribution, i.e., we do not contradict Fig.~\ref{fig:joint_massfrac} that shows that this mass is unconstrained by our data.}
    \label{Tab:mass_lims}
    %make it clear that this is just for 10^-24, reference figure 18 for mass dependence
  %  \end{centering}
\end{table}   

\vspace{0.5in}

\section{Discussion}
\label{sec:Discussion}

%Hubble measurements of the UVLF probe a range of redshifts and scales that have not been probed by measurements of large scale clustering, the Lyman-$\alpha$ forest, or the CMB. Since the impact of axions on structure formation is enhanced at higher redshifts and smaller scales, this allows us to place tight constraints on axion mass and fraction without needing to model the complexities of axion astrophysics. 

%key points

%P1
%is ST good enough? we find that axions impact different models the same way
%why we can neglect axion halo pressure and axion astrophysics effect, might need it more in the future with smaller-scale higher-redshift measurements from JWST, use Schrodinger-Poisson simulations to accurately capture axion + baryonic effects
%schive, mocz, lague
%not very many detailed mixed fuzzy simulations to check with
%could map to other WDM, etc, since we only care about linear MPS

\subsection{Halo mass function model}

In this work, we use the Sheth-Tormen model of ellipsoidal collapse \citep{Sheth:1999mn} to compute the HMF, using parameters fit to pure-CDM simulations, as described in Sec.~\ref{sec:HMF}. This model has been used in the past (such as by \cite{Vogt:2022bwy}) to model the formation of halos in mixed-axion cosmologies, and compared to simulations in pure axion cosmologies in \cite{Dentler:2021zij}. We show in Appendix~\ref{sec:AppC} that alternative forms for the mass function (such as those proposed in \cite{Reed:2006rw}) are impacted by the linear axion power spectrum in almost identical ways. We find that the impact of axion halo pressure on the HMF is limited to either low halo masses beyond the sensitivity of the UVLF or there is a negligible suppression relative to the impact of the linear matter power spectrum (see Fig.~\ref{fig:HMF_axions}). Since the impact of axions on the HMF from the primordial linear matter power spectrum is found to be largely independent of the choice of halo formation model and the effects of axion halo pressure are negligible for the axion mass and fraction ranges that we consider, we conclude that our estimates of the HMF are sufficiently accurate compared to the data uncertainties.

Future inferences of the HMF, including using JWST UVLF to probe higher redshifts and smaller scales, will become more sensitive to the small-scale effects of axion halo pressure, necessitating more robust modeling of the impact of axion physics on halo formation, e.g., the role of axion quantum pressure in the collapse of overdensities. This would require comprehensive Schrodinger-Poisson simulations in order to capture the combination of axion, baryonic and CDM physics involved in halo formation in mixed-axion cosmologies, such as those conducted by \cite{Schwabe:2020eac, Huang:2022ffc, Mocz:2023adf, Lague:2023wes}. A consequence of our findings, that the impacts of small-scale axion physics on halo formation are negligible given current data, is that our methods can be used to test other DM models with a suppressed linear power spectrum (such as warm \citep[e.g.,][]{Viel:2013fqw, Rudakovskyi:2021jyf, Maio:2022lzg, Liu:2024edl} or interacting \citep[e.g.,][]{Rogers:2021byl} DM).

%P2
%mean UV luminosity model (not star formation model)
%assume gaussian scatter to account for schotasitcsicity, accoutn for bursty star formation, etc.
%comparison to simulations with realistic models of star formation could improve the model
%we are confident that the model is accurate enough given the precision of the data, 
%flexible parametric model that is shown to capture the features of simulations and fit current Hubble data well
%we test different models, degree of consistency, but we choose conservative bounds
%kanan et al
%big hydro sims, like Illustris TNG, Bahamas, cosmoOWLS, look into emulator literature
%not tuned to specific simulation, but flexible enough to capture different

\subsection{UV luminosity model}
\label{sec:UVLF_discussion}

Our results depend on the mean UV luminosity of halos as described in Sec.~\ref{sec:UVLF}. The flexible parametric model that we use is shown in \cite{Sabti:2021unj} to capture the results of hydrodynamical simulations such as Illustris TNG \citep{IllustrisTNG_JWST} and to fit Hubble UVLF data well under the \(\Lambda\)CDM model. In Appendix~\ref{sec:AppB}, we also test an alternative model of UV magnitude, informed by empirical measurements of low-$z$ galaxies, and we find that our chosen model gives constraints that are either consistent or conservative in comparison. We then relate the HMF and UVLF using a Gaussian scatter of UV magnitudes around this mean relation, in order to account for stochasticity from galaxy composition and bursty star formation \citep[e.g.,][]{2023ApJ...955L..35S, Heather:2024elk}. Importantly, when searching for the signature of axions, we marginalize over uncertainty in the stellar mass to halo mass relation in both its scale and time dependence, covering the uncertainties seen in simulations. Remarkably, despite this marginalization, we still attain powerful sensitivity to axions as the effect of axions is shown to be often qualitatively distinct.

We consider the impact of axion astrophysics on the UV luminosity of low-mass halos, where halos below the halo Jeans critical mass will lack an axion DM component and thus have a higher baryon-to-DM ratio, enhancing UV luminosity. However, we show in Fig.~\ref{fig:UVLF_mass} that the impacts of axion astrophysics on the UVLF are negligible compared to the axion cosmology through the linear MPS (in particular, when accounting for the cancelling effect from when axion halo pressure suppresses the formation of low-mass halos). We therefore neglect these axion impacts on mean UV luminosity.

Comparison to simulations with even more realistic models of star formation and radiative transfer at high redshifts can further improve this model of UV luminosity \citep[such as those done by][]{McCarthy:2016mry, IllustrisTNG_JWST, Kannan:2021xoz, Maio:2022lzg,2023ApJ...955L..35S, Shen:2023lsf}. This improved modeling can also be informed by future multi-wavelength studies of high-$z$ galaxies using HST, JWST and the Roman Space Telescope \citep{Montes:2023nvh}, characterizing the star formation process at these times \citep[as discussed in, e.g.,][]{Munoz:2023cup, Fujimoto:2023orx, Sabti:2023xwo, Heather:2024elk}. \rev{For example, \cite{Song:2016} suggests a $z$-dependence of the low halo mass slope of the stellar to halo mass relation $\alpha_\star$ which can be incorporated into our model for added flexibility.} However, both observations and simulations currently have high uncertainties at redshifts \(z > 10\) beyond what we probe with HST data.

% More detailed future results from JWST characterizing the star formation process at these high redshifts (such as those done by \cite{Fujimoto:2023orx}) could help us inform our astrophysical model of UV luminosities, as would more detailed simulations of galaxy formation informed by these high-$z$ observations (such as those done by \cite{Kannan:2021xoz}). A more detailed model of star formation at high redshifts could motivate alternative functional forms for our model, or suggest well-motivated priors on the model parameters. This information could help us constrain the nuisance parameters of our model, leading to tighter and more physically-motivated constraints on the axion cosmology.

%P3
%Webb and additional high-z data
%story is that JWST is giving more structure than expected at high redshifts compared to whart you;d expect from just Hubble
%fitting just JWST is overfit, but no constraints on 
%clarify around discrepency between datasets
%we allow redshift-dependance, which allows us to fit both hubble and webb
%make it clear that we need other measurements at other wavelengths to better constrain star formation model
%current simulations have uncertainty at high redshift, newer measurements will provice additional data (not priors) on astrophysical element of model
%nothing discrepant with LCDM
%talk about extreme axions here?
%some photometric datapoints that seem to push more on predictions from Hubble + CDM, bring in extreme axions in this context

\begin{figure}
    \centering
    \includegraphics[trim={0.3cm 0.3cm 0.2cm 0.2cm }, clip,width=\linewidth]{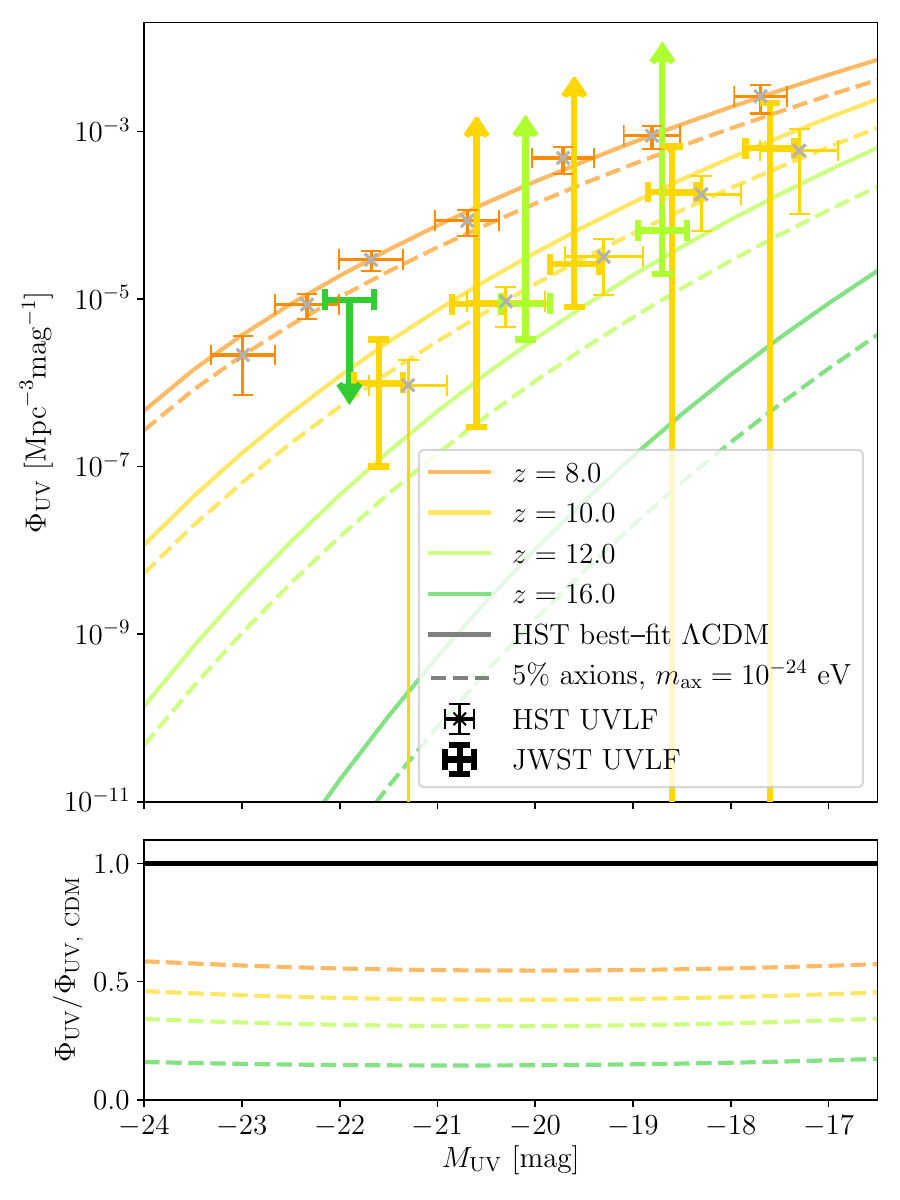}
    \caption{JWST estimates of the UVLF at redshifts $z=10,12,16$ in thick error bars from \cite{JWST_UVLF}. Down arrows represent upper limits, while up arrows represent lower limits with uncertainties. We also show (in thin error bars) HST estimates of the UVLF at redshifts $z=8,10$. The solid curves show the best-fit $\Lambda$CDM cosmology given HST data, while the dashed lines are with 5\% axion DM and $m_\mathrm{ax} = 10^{-24}$ eV. The bottom panel shows the ratio of the UVLF in the presence of axions relative to the \(\Lambda\)CDM case. The impacts of axions on the UVLF are proportionally higher at higher redshift, but current spectroscopically-confirmed JWST samples do not have the required sensitivity to distinguish this effect after marginalization over astrophysical uncertainties.}
    \label{fig:UVLF_JWST}
    %change colors to be consistent with z, go beyond yellow to green?
\end{figure}

\subsection{Preliminary results from James Webb Space Telescope luminosities}

There is much debate in the community regarding the unexpected detection of numerous massive high-redshift (\(z > 10\)) galaxies observed by JWST \citep{Yan2023}. Early results claimed to be inconsistent with HST predictions, suggesting that our current models of early star and galaxy formation may be incomplete \citep[see e.g.][]{Bouwens2023, Chemerynska2023}. If robust, these observations have the potential to place tighter constraints on axion DM physics or to force us to reevaluate our models of halo formation and UV luminosity at higher redshifts.

In order to test whether these preliminary JWST results are in tension with our model, we consider a spectroscopically-confirmed sample of 25 JWST UV sources presented in Table 1 of \cite{JWST_UVLF}. We restrict ourselves to only spectroscopically-confirmed sources as photometric samples are known to be contaminated by low-\(z\) interlopers. It is only with mid- and far-infrared spectroscopy that redshifts can be definitively determined. Indeed, many of the initial high-\(z\) sources that were claimed to be in tension with \(\Lambda\)CDM cosmology were later found to be interlopers and determining the true redshift is still an active area of research \citep{Zavala2023, Fujimoto:2023orx, Heather:2024elk}. The sources to which we restrict our analysis have spectroscopic redshifts from $8.61 < z < 13.20$, allowing \cite{JWST_UVLF} to compute the UVLF at $z = 9$ and $z = 10$. They also compute lower bounds at $z = 12$, and upper bounds at $z = 16$ (from non-detection in their field). Fig.~\ref{fig:UVLF_JWST} shows the current JWST estimates of the UVLF, along with estimates of the UVLF from our model fit to HST. The JWST data are consistent with both the HST data at comparable redshifts and the best-fit UVLF model. We show in the bottom panel that the effect of axions increases with redshift.

%there is also a photometric sample, but effected by leakage from low-z interlopers
%tension is higher in this sample

 \begin{figure}
    \centering
    \includegraphics[trim={0.2cm 0.2cm 1.0cm 1.0cm }, clip,width=1.\linewidth]{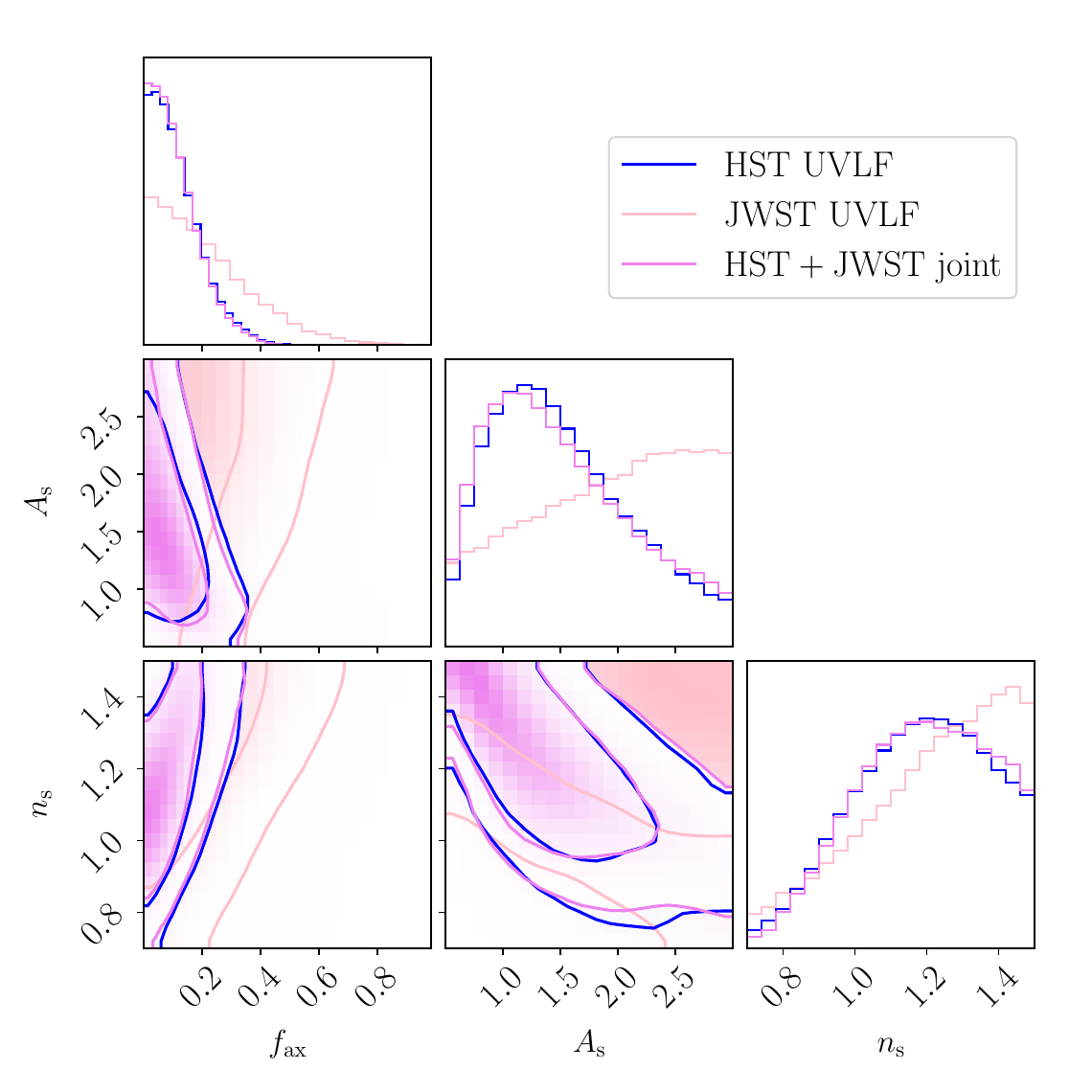}
    \caption{As Fig.~\ref{fig:corner_joint_cosmo}, where \textcolor{HWblue}{blue} contours show constraints from HST UVLF \citep{HST_UVLF}, while \textcolor{HWpink}{pink} contours are from a sample of 25 spectroscopically-confirmed galaxies from JWST \citep{JWST_UVLF}. The \textcolor{HWmagenta}{magenta} contours represent joint constraints from both HST and JWST. \(A_\mathrm{s}\) is given in units of \(10^{-9}\). While JWST alone sets a limit on the axion fraction, the joint constraints are not meaningfully different than the HST-only case.}
    \label{fig:corner_JWST}
\end{figure}

As with the HST UVLF (Sec.~\ref{sec:HST_UVLF}), we combine the JWST UVLF with the Pantheon supernovae likelihood, along with BBN priors on $\omega_\mathrm{b}$ and a fixed value of $100 \theta_\mathrm{s} = 1.040827$. We show marginalized posteriors given JWST data in Fig.~\ref{fig:corner_JWST}. The pale pink contours show constraints on $m_\mathrm{ax} = 10^{-24}$ eV axions from JWST data alone, while blue contours are from HST UVLF alone. Despite only 25 sources, compared to HST's 24,000 sources, JWST already puts a comparable limit on the axion fraction:
\begin{eqnarray}
f_\mathrm{ax} [m_\mathrm{ax}=10^{-24} \mathrm{eV}] &<& 0.52 \,\,\mathrm{(95\% \ C.L.)} \nonumber \\ &&[\mathrm{JWST}~\mathrm{UVLF}]. \nonumber\\
\end{eqnarray} 
This limit is tighter than the \textit{Planck} CMB limit of 0.93 (see Eq.~\eqref{eq:result_Planck}), but looser than the HST UVLF limit of 0.28 (see Eq.~\eqref{eq:result_HST}). %The flexible parametric model of UV luminosity is almost entirely unconstrained by JWST data alone, owing to the high number of model parameters relative to the amount of UVLF data. 
The magenta contours in Fig.~\ref{fig:corner_JWST} indicate joint constraints from HST and JWST UVLF showing that JWST data does not meaningfully improve axion constraints set by HST at this time. Our model (a flexible double power law given in Eq.~\eqref{eq:double_power} with a redshift-dependant amplitude $\epsilon_\star$ and pivot mass $M_c$) is consistent with both HST and spectroscopic JWST results. These JWST observations do not present any discrepancy with $\Lambda$CDM.

Any tests of discrepancy at $z>10$ will require additional spectroscopic data. The effect of axions (dashed lines in Fig.~\ref{fig:UVLF_JWST}) is to suppress the UVLF at all redshifts (from ~40\% at $z=8$ to ~90\% at $z=16$). If future JWST measurements prove an excess of sources over a larger and more accurately determined volume factor, this will further constrain the axion fraction, particularly since the relative impact of axions is greater at higher redshifts (see the lower panel of Fig.~\ref{fig:UVLF_JWST}). Detections of axion suppression will require a large volume factor in order to measure a deficiency of high-$z$ UV sources, while limiting the axion fraction could be achieved with a smaller volume factor. We leave a thorough computation of JWST UVLF forecasts on axions for future study. This approach is also complementary to ongoing efforts to use 21-cm measurements to detect axions at similar scales and redshifts \citep{HERA:2021noe, Hotinli:2021vxg, Flitter:2022pzf}.

Observations of abundant structure at high redshifts could imply the presence of extreme axions (i.e. axions with a high starting field angle, resulting in the growth of field perturbations and an overabundance of structure on certain scales), as described in \cite{Leong:2018opi} and \cite{Arvanitaki:2019rax}. Although we do not consider the impacts of extreme axions in this paper, such a comparison could be performed using efficient extreme-axion computations, such as those developed by \cite{Winch:2023qzl} for the linear MPS. %We leave this for the assessment of the impacts of extreme axion physics on the UVLF is left for future work
%p4
%small-scale structure, talk about spoon
%list of other ways of measuring small structure

\subsection{Small-scale structure}

Our work also presents evidence for a new observable consequence of mixed-axion DM models: an overabundance of low-mass halos at late times (as shown in Fig.~\ref{fig:HMF_m25_z}). Our modeling calculates that this effect is maximized for axion fractions between $5-10\%$. This effect is due to the axion component suppressing primordial linear structure on small scales, delaying the hierarchical growth of halos and resulting in an overabundance of low-mass halos that have not yet merged into higher mass ones at late times. The halo masses ($M_h \lesssim 10^9 M_\odot$)  and redshifts ($z \lesssim 3$) relevant for this effect are both too low to be measured using the UVLF in this paper. But there is an opportunity to search for these effects in late-time small-scale galaxy structure (e.g., using stellar streams \citep{Banik:2019smi}, satellite galaxies \citep{Nadler:2021dft} or strong gravitational lensing \citep{Shevchuk:2023ccb}). A more detailed treatment of axion astrophysics would be necessary as the effects of axion wave physics would have a significant impact on hierarchical merger histories on small scales. A more thorough computation is left for future study.

\section{Conclusions}
\label{sec:Conclusions}

%first demonstration of small-scale high-redshift constraints on the nature of DM
%find that high-z UVLF is consistent with LCDM
%can still find conservative constraints on BSM DM physics
%
%fill the window in axion mass, without any assumptions about soliton profile modeling

%find consistency with the preference for axions foudn in Rogers and Poulin
%axions also alleviate
%consistent with the mixed-axion contributions that are invoked to alleviate small-scale discrepency (rogers and poulin, and S8 paper)
%early-time consistency check of solutions late-time problems

%no tension between JWST and HST if you allow for uncertainty in redshift-dependance of star formation model

%paragraph

The galaxy UV luminosity function  is a powerful probe of structure on small scales (\(0.5\,\mathrm{Mpc}^{-1} < k < 10\,\mathrm{Mpc}^{-1}\)) and high redshifts (\(4 \leq z \leq 16\)). We demonstrate, for the first time, by using a flexible parametric model relating UV luminosity and halo mass, the use of the UVLF to set world-leading limits on fundamental physics. Our main conclusions are:

%Although it does depend on astrophysical models of stellar mass and star formation, which are poorly constrained at early times, we can assume generic models with uniform priors on astrophysical parameters and still impose novel cosmological constraints on ultralight axion DM.

\begin{enumerate}
     \item \textbf{The UVLF depends primarily on the mixed-axion linear matter power spectrum} and is largely independent of small-scale axion astrophysics (see Sections \ref{sec:HMF} and \ref{sec:UVLF} and Appendices \ref{sec:AppB} and \ref{sec:AppC}).
     %In sections \ref{sec:HMF} and \ref{sec:UVLF}, we investigate the impact of axion halo pressure on the HMF and axion astrophysics on the UVLF, but conclude that these effects are negligible when compared to the effect of axion cosmology on the linear MPS. We also find that our results are largely resilient to alternative choices for the halo formation or mean UV luminosity models, as shown in Appendices \ref{sec:AppB} and \ref{sec:AppC}.
     
    \item \textbf{We set new cosmological constraints on mixed-axion cosmologies} by computing the impacts of axions on the HMF and the UVLF, while robustly marginalizing over uncertainty in the relation between UV luminosity and halo mass.
    %We compare these predictions of the UVLF to measurements using the Hubble Space Telescope (HST) at a range of redshifts ($4 < z < 10$), varying the axion mass and DM fraction, and marginalizing over all relevant cosmological and astrophysical parameters. When combined with precise measurements of large-scale fluctuations from \textit{Planck} CMB, we are able to achieve constraints on axion DM that surpass current cosmological probes (including CMB and Ly-$\alpha$ forest experiments). In addition, the UVLF utilizes a range of redshifts previously unexplored by any other axion probe, providing a powerful consistency check of low-$z$ measurements of axion DM, such as the reported preference for mixed-axion DM reported in \cite{Rogers:2023upm}. 
    Joint HST UVLF (for \(4 \leq z \leq 10\)) and \textit{Planck} CMB likelihoods rule out axions with $\log( m_\textrm{ax}/\textrm{eV}) < -21.6$ each as 100\% of the DM and limit axions with $\log( m_\textrm{ax}/\textrm{eV}) < -23$ to constitute $\leq 22\%$ of the DM (both limits at $95\%$ credibility; detailed constraints given in Table \ref{Tab:mass_lims}). Our results are the first joint constraints combining \textit{Planck} CMB with small-scale measurements, constraining axion DM on a wide range of cosmological scales and bridging the mass gap in axion constraints shown in Fig.~\ref{fig:joint_massfrac_compare}. 
    
    \item \textbf{Current JWST spectroscopic measurements of the UVLF are consistent with $\Lambda$CDM} when computed using a flexible parametric model of UV luminosity with redshift-dependant amplitude and pivot mass. As shown in Fig.~\ref{fig:corner_JWST}, combining HST estimates of the UVLF with spectroscopic JWST data from \cite{JWST_UVLF} does not meaningfully improve our constraints over HST alone. However, we do find in Fig.~\ref{fig:HMF_m25_z} that the impact of axions on the HMF and UVLF is more pronounced at higher redshifts, implying that future high-$z$ estimates of the UVLF from JWST have the potential to improve our constraints significantly. More detailed multi-wavelength measurements of high-$z$ galaxies with JWST will also help to refine our picture of star formation at early times, informing our assumptions about how UV luminosity relates to halo mass at high redshifts (\(z > 10\)). As we enter a new era of high-redshift galaxy science with JWST, the UVLF will continue to improve as a powerful probe of the nature of dark matter.

\end{enumerate}

%#We compute the impact of mixed-axion models of DM on the HMF and, by extension, the UVLF. 
%We assume a generic astrophysical model of star formation to compute the UVLF, and margianlize over all astrophysical model parameters. 
  
%These constraints are completely independent of Ly-$\alpha$ noise, systematics and astrophysical modeling, providing a powerful complementary probe of axion physics. 

%we find that axions impact things more on high-z, therefore...
%measurements at other wavelengths to help our model
\begin{figure*}[htbp!]
    \centering
    \includegraphics[trim={2.1cm 0.2cm 3cm 0.3cm }, clip,width=\linewidth]{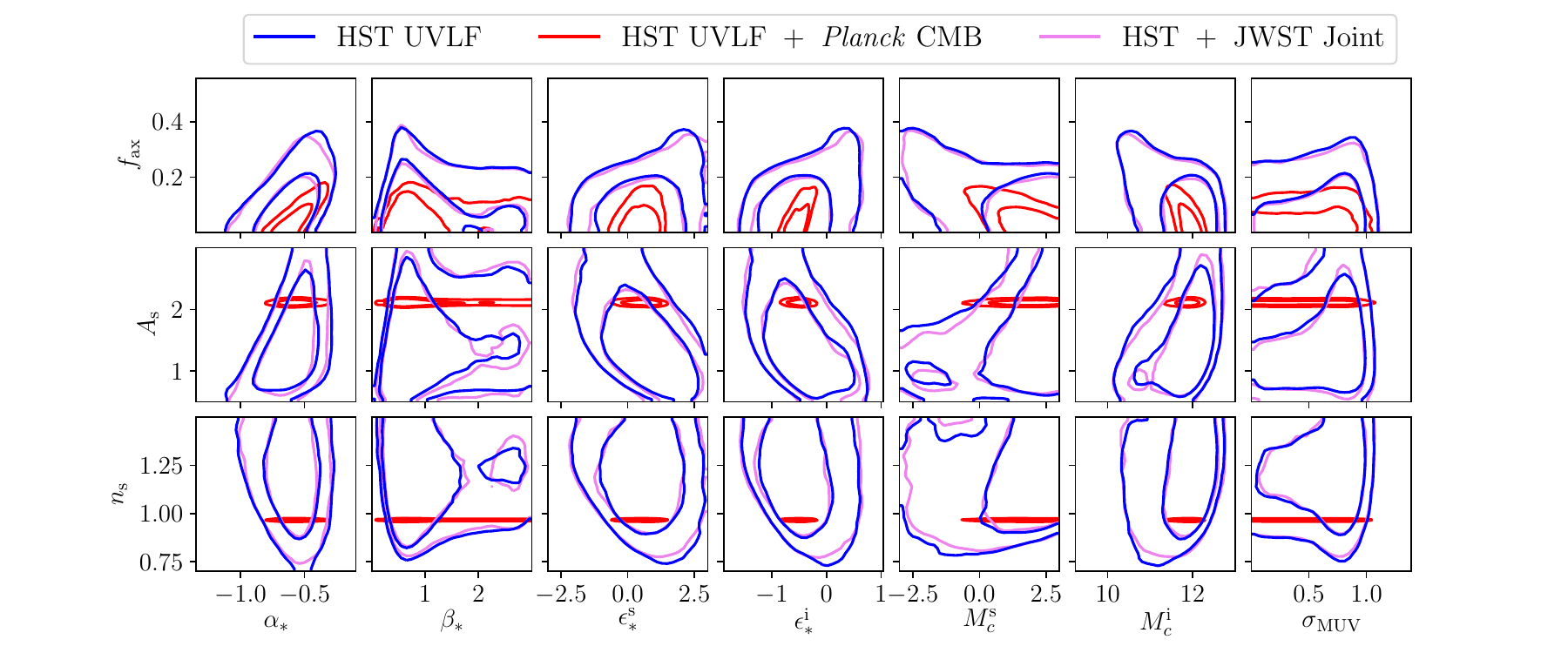}
    \caption{Marginalized posterior on cosmological parameters (along the vertical axis: $f_\textrm{ax} = \Omega_\mathrm{ax}/\Omega_\mathrm{DM}$ for axions with a mass of $10^{-24}$ eV, $A_\mathrm{s}$ and $n_\mathrm{s}$) and astrophysical parameters (along the horizontal axis, as defined in Eq.~\eqref{eq:double_power}). The \textcolor{HWblue}{blue} contours are from the HST UVLF, while the \textcolor{HWred}{red} contours are from the HST UVLF plus \textit{Planck} CMB. Imposing the \textit{Planck} likelihood in addition to HST UVLF breaks degeneracies with \(A_\mathrm{s}\) and \(n_\mathrm{s}\) and therefore tightens constraints on the astrophysical parameters of the UVLF model. We also show in \textcolor{HWmagenta}{magenta} the joint HST and JWST UVLF posterior, which is nearly identical to the HST-only case. Inner and outer contours respectively show the 68\% and 95\% credible regions of the marginalized posterior. \(A_\mathrm{s}\) is given in units of \(10^{-9}\). Table~\ref{Tab:constraint_combos} describes the full datasets that we use.}
    \label{fig:corner_joint_astrocross}
\end{figure*}

\section{Acknowledgements}

We thank Daniel Grin, Rahul Kannan, Adam Lidz, Julian Mu\~noz, Vivian Poulin, Nashwan Sabti and Jackson Sipple for their helpful advice and discussion. H.~W. acknowledges the support of the Natural Sciences and Engineering Research Council of Canada (NSERC) Canadian Graduate Scholarships - Doctoral program. The Dunlap Institute is funded through an endowment established by the David Dunlap family and the University of Toronto. The authors at the University of Toronto acknowledge that the land on which the University of Toronto is built is the traditional territory of the Haudenosaunee, and most recently, the territory of the Mississaugas of the New Credit First Nation. They are grateful to have the opportunity to work in the community, on this territory.
R.~H. acknowledges support from CIFAR, the Azrieli and Alfred. P. Sloan foundations and is supported by the Natural Sciences and Engineering Research Council of Canada Discovery Grant Program and the Connaught Fund. D.~J.~E.~M. is supported by an Ernest Rutherford Fellowship from the Science and Technologies Facilities Council (ST/T004037/1) and by a Leverhulme Trust Research Project (RPG-2022-145).

 %\begin{figure*}
 %   \centering
 %   \includegraphics[trim={0cm 0cm 1cm 1cm }, clip,width=1.05\linewidth]{final_cornerplot_joint_astro.pdf}
 %   \caption{Constraints on the axion fraction ($ f_\textrm{ax} = \Omega_{ax}/\Omega_{DM}$) for axions with a mass of $10^{-24}$ eV, as well as $A_s$, $n_s$, and all astrophysical parameters that go into computations of the UVLF, as defined in Eq. \eqref{eq:double_power}. The blue contours are from the HST UVLF data shown in Figure \ref{fig:UVLF_fiducial_z} (plus Pantheon Sne data and a BBN prior on $\Omega_b$), while the red contours are from all of those likelihoods plus \textit{Planck} CMB data (TT, TE, EE, and lensing). We can see how imposing the \textit{Planck} likelihood in addition to HST UVLF anchors the cosmological model, and therefore tightens the constraints on the astrophysical parameters of our star formation model.}
%    \label{fig:corner_joint_astro}
%\end{figure*} 

\begin{figure*}
    \centering
    \includegraphics[trim={0cm 0cm 1cm 1cm }, clip,width=\linewidth]{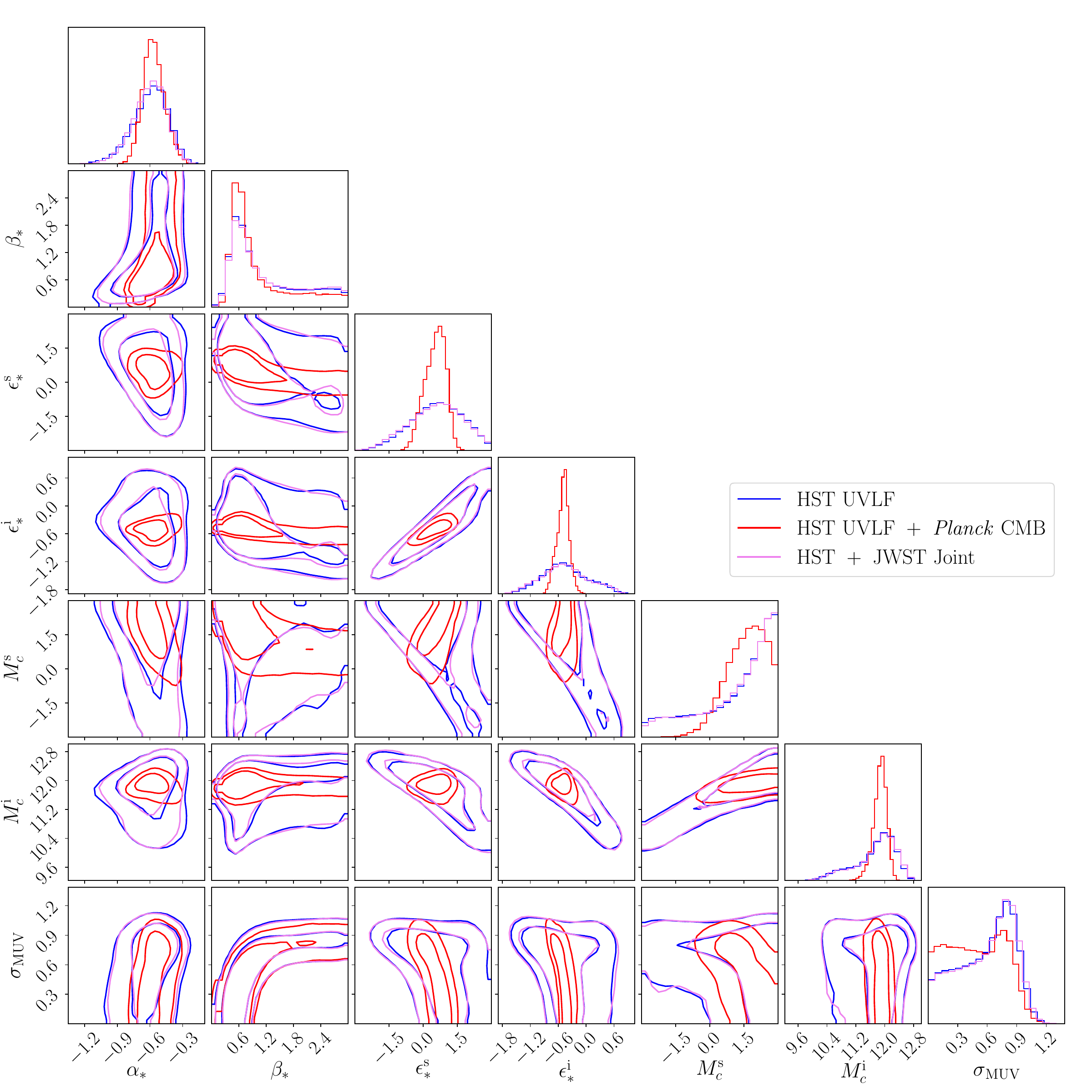}
    \caption{Marginalized posterior on astrophysical parameters (as defined in Eq.~\eqref{eq:double_power}), with the same data combinations as in Fig.~\ref{fig:corner_joint_astrocross}. Inner and outer contours respectively show the 68\% and 95\% credible regions of the marginalized posterior. Table~\ref{Tab:constraint_combos} describes the full datasets that we use.}
    \label{fig:corner_joint_astro}
\end{figure*}

\appendix

\section{Constraints on astrophysical parameters}
\label{sec:AppA}

%As discussed in Sec.~\ref{sec:Results} and shown in Fig.~\ref{fig:corner_joint_cosmo}, by combining HST measurements of the UVLF with \textit{Planck} CMB measurements of the underlying cosmology, we can achieve joint constraints on the axion fraction that surpass either experiment individually. The UVLF does not allow large values of the axion fraction, while being relatively insensitive to either the amplitude of scalar fluctuations $A_s$ or the scalar spectral index $n_s$.  
%This is primarily due to the differing estimates of the scalar amplitude and spectral index, $A_s$ and $n_s$. 
%Anchoring the $A_s$ and $n_s$ values using the \textit{Planck} likelihood effectively fixes the amount of small-scale power. The higher values for the HST UVLF can no longer be explained by increased $n_s$ values, so instead they must be explained through compensation in the astrophysical nuisance parameters describing the UV luminosity. 

Figure \ref{fig:corner_joint_astrocross} shows constraints on axion fraction, cosmological and UVLF parameters from Table \ref{Tab:param_priors} for an axion mass of $10^{-24}$ eV. The blue contours are given HST UVLF alone, while red contours are given joint HST UVLF and \textit{Planck} CMB likelihoods. 
%Imposing the \textit{Planck} CMB likelihood in addition to the HST UVLF shifts the posterior on $\epsilon_\star^i$ and $\epsilon_\star^s$ to higher preferred values, since $\epsilon_\star$ scales the overall amplitude of the stellar mass function. \textit{Planck} anchors on $n_s$ prefer lower small-scale power, so the model compensates by increasing stellar mass for halos of a fixed total mass.  
We also show in magenta the constraints given joint HST + JWST UVLF estimates, which are identical to the HST-only results. Fig.~\ref{fig:corner_joint_astro} shows the correlations between all UVLF model parameters, for HST UVLF only, HST UVLF + \textit{Planck} CMB and HST + JWST UVLF. The addition of \textit{Planck} data tightens the constraint on some UVLF parameters by breaking degeneracy with cosmological parameters.

\begin{figure*}
    \centering
    \includegraphics[width=\linewidth]{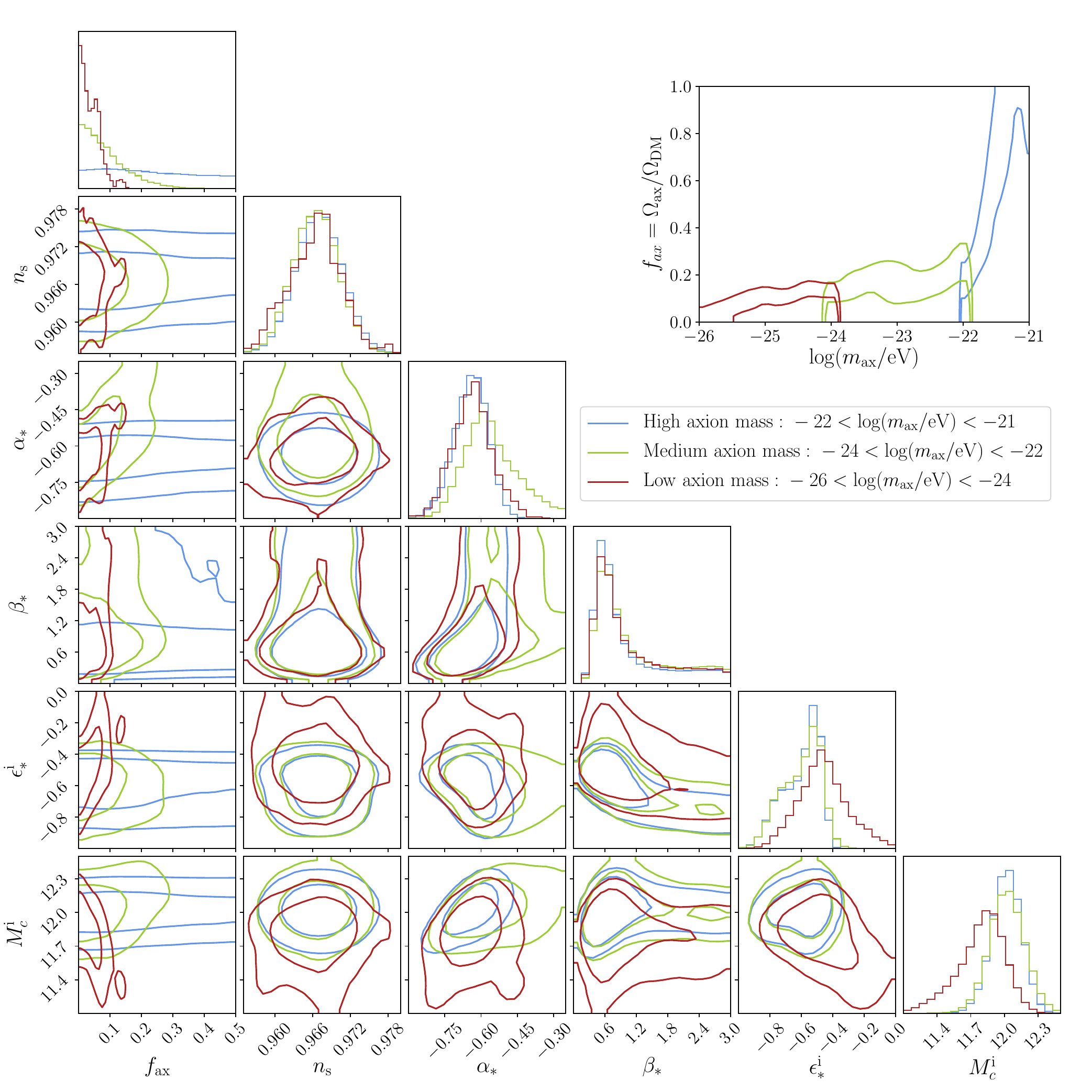}
    \caption{Constraints on the axion fraction and some cosmological and astrophysical parameters from joint HST UVLF and \textit{Planck} CMB likelihoods, divided into three axion mass bins as shown in the upper-right subplot. Constraints on high mass axions ($-22 < \log( m_\textrm{ax}/\mathrm{eV}) < -21$) are shown in blue, medium mass axions ($-24 < \log( m_\textrm{ax}/\mathrm{eV}) < -22$) are shown in green, and low mass axions ($-26 < \log( m_\textrm{ax}/\mathrm{eV}) < -24$) are shown in red. Inner and outer contours respectively show the 68\% and 95\% credible regions of the marginalized posterior.}
    \label{fig:corner_joint_massbins}
\end{figure*}

In Fig.~\ref{fig:corner_joint_massbins}, we show the joint HST UVLF and \textit{Planck} CMB constraints on axions for axion masses in three different ranges.
%add sentance about high masses
%the degeneracy with axions and astrophysical parameters varies with mass, so depending on which aspects of the astrophysical model we get measure
The degeneracy direction of axion fraction and certain astrophysical parameters changes with axion mass bin. Medium mass axions ($-24 < \log( m_\textrm{ax}/\mathrm{eV}) < -22$, in green) have a positive degeneracy with $\alpha_\star$, since their impact on the UVLF primarily occurs at faint luminosities. Low-mass axions ($-26 < \log( m_\textrm{ax}/\mathrm{eV}) < -24$, in red) have a positive degeneracy with $\epsilon_\star^i$, since they suppress the UVLF on a wider range of scales (see Fig.~\ref{fig:UVLF_mass}). 
It would be powerful to identify complementary datasets that break these degeneracies, bearing in mind that these degeneracies vary with axion mass.
%This demonstrates how constraining the different features of the stellar mass function (such as $\alpha_\star$ or $\epsilon_\star$, given in \eqref{eq:double_power}) using data at a range of redshifts could allow us to further tighten our constraints on axions in certain mass bins by constraining degenerate astrophysical parameters.

\vspace{2cm}

\section{An alternative model of the UV luminosity function}
\label{sec:AppB}

 \begin{figure*}
    \centering
    \includegraphics[trim={0cm 0cm 1cm 1cm }, clip,width=0.9\linewidth]{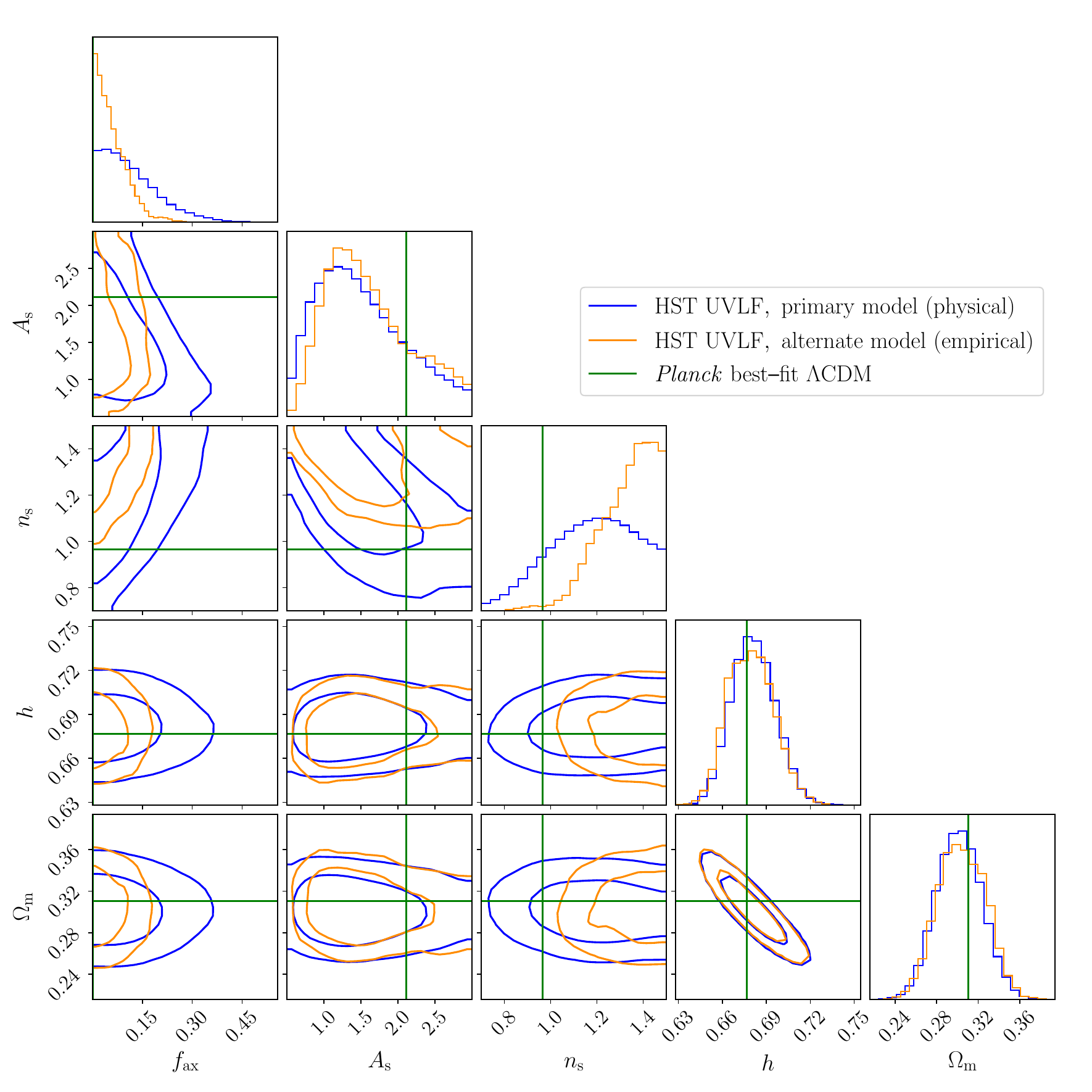}
    \caption{Marginalized posterior on cosmological parameters and axion fraction for $m_\textrm{ax} = 10^{-24}$ eV using the HST UVLF with two different models of the UVLF. The primary model (in blue) corresponds to Model II in \cite{Sabti:2021xvh}, which relates stellar mass and UV luminosity through a physically-motivated argument, while the alternative model (in orange) corresponds to Model III in \cite{Sabti:2021xvh}, which relates stellar mass and UV luminosity through empirical estimates at $4 < z < 10$ using HST infrared observations \citep{Stefanon:2021}. The green lines show the best-fit $\Lambda$CDM cosmological parameters given \textit{Planck} CMB.}
    \label{fig:corner_models}
\end{figure*}

In Section \ref{sec:UVLF}, we discuss our method for relating halo mass to UV luminosity, which follows Method II in \cite{Sabti:2021xvh}. Here, we discuss their Model III and use it to compute constraints on mixed axion cosmologies, demonstrating consistency with Model II. In this alternative model, we relate $M_h$ and $\overline{M_\star}$ using the same double power law as in Eq.~\eqref{eq:double_power} and the same parameterization of $\alpha_\star, \beta_\star, \epsilon_\star^s, \epsilon_\star^i, M_c^s, M_c^i$. However, we then use an empirically-determined relation between $M_\star$ and $M_\mathrm{UV}$, which is derived from near-infrared observations of galaxies, presented in \cite{Song:2016, Stefanon:2021}. The stellar mass is calculated from stellar synthesis population models compared to the multi-band observations of HST galaxies and this is then compared to the rest-frame UV luminosity. This relation is calculated for galaxies from $z=4$ to $z=10$ and is estimated to be accurate to within a factor of two \citep{Sabti:2021xvh}.

In Fig.~\ref{fig:corner_models}, we show constraints on cosmological parameters using both the primary (Model II in \cite{Sabti:2021unj}) and alternative (Model III in \cite{Sabti:2021unj}) methods of computing the UVLF. We show in green the \textit{Planck} $\Lambda$CDM best-fit parameter values for comparison. Both methods give largely consistent constraints on cosmological parameters. However, the alternative method gives slightly tighter constraints on the axion fraction, while also preferring higher values of $n_\mathrm{s}$ that are mildly inconsistent with \textit{Planck} $\Lambda$CDM values. Therefore, we use the primary model of UV luminosity for our fiducial study, due to its greater consistency with \textit{Planck} and more conservative limit on the axion fraction.

\section{An alternative model of the halo mass function}
\label{sec:AppC}

In Section \ref{sec:HMF}, we use the Sheth-Tormen mass function \citep{Sheth:1999mn} when computing the HMF, as done in \cite{Sabti:2021unj}. However, alternative mass functions have been proposed to describe halo formation. Here, we consider the alternative Reed mass function \citep{Reed:2006rw} and compare the effects of axions on the HMF using both methods. In the Reed model, Eq.~\eqref{eq:ShethTormen} is replaced by:
\begin{equation}
    f_\mathrm{Rd}(\sigma_{M_\mathrm{h}}) = A_\mathrm{Rd}\sqrt{\frac{2 a_\mathrm{Rd}}{\pi}}\left[1 + \left(\frac{\sigma^2_{M_\mathrm{h}}}{a_\mathrm{Rd}\delta^2_\mathrm{Rd}}\right)^{p_\mathrm{Rd}} + 0.2\exp\left(-\frac{(\ln\sigma^{-1}_{M_\mathrm{h}} - 0.4)^2}{2(0.6)^2}\right)\right] \frac{\delta_\mathrm{Rd}}{\sigma_{M_\mathrm{h}}} \exp\left(-\frac{c_\mathrm{Rd}a_\mathrm{Rd}\delta_\mathrm{Rd}^2}{2 \sigma^2_{M_\mathrm{h}}}\right),
\end{equation}
where $A_\mathrm{Rd} = 0.3235$, $a_\mathrm{Rd} = 0.707$, $p_\mathrm{Rd} = 0.3$, $c_\mathrm{Rd} = 1.081$ and $\delta_\mathrm{Rd} = 1.686$ \citep{Reed:2006rw,Sabti:2021unj}.

\begin{figure}[htbp!]
    \centering
    \includegraphics[trim={0cm 0cm 0cm 0cm }, clip,width=0.55\linewidth]{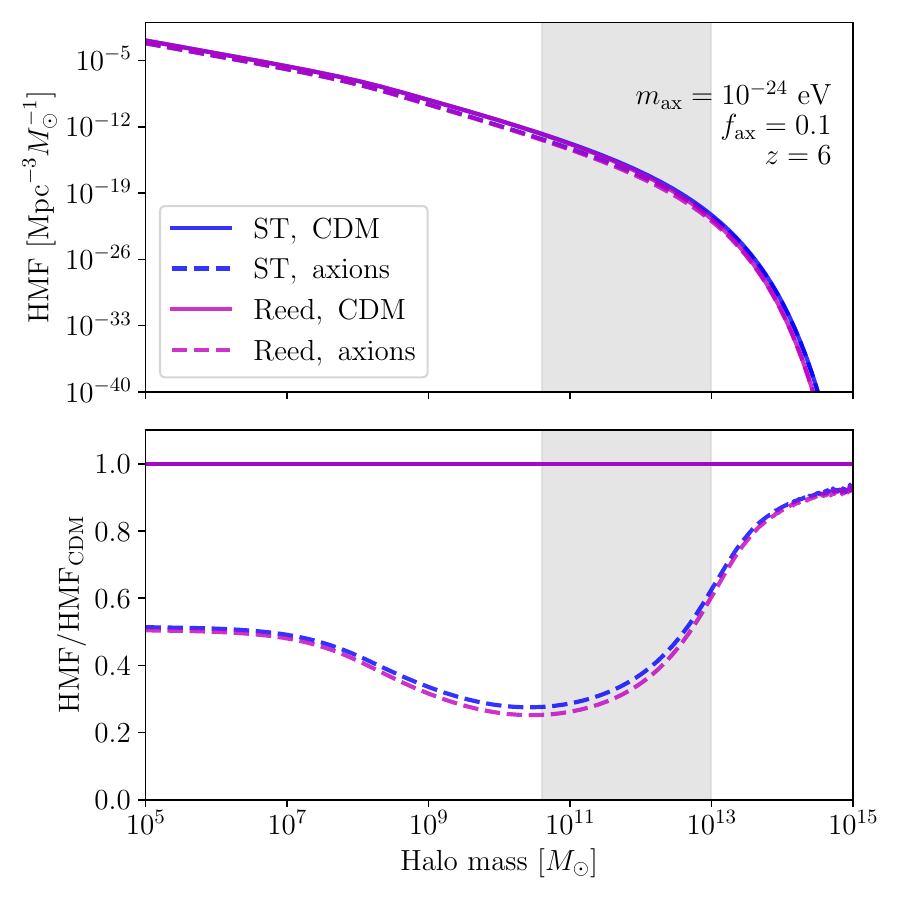}
    \caption{The HMF computed using both the Sheth-Tormen (ST, in blue) and Reed (in magenta) multiplicity functions. The solid lines are for a fiducial CDM cosmology, while the dashed lines show axions with $m_\mathrm{ax} = 10^{-24}$ eV as 10\% of the DM. The HMF given each multiplicity function is impacted by axions in almost identical ways, meaning that our final axion constraints are independent of the choice between the two.}
    \label{fig:HMF_mass_func}
\end{figure}

\cite{Sabti:2021unj} compares astrophysical and cosmological impacts on the UVLF given the two mass functions. They find that, while the high-$z$ estimates of the UVLF are slightly changed by the Reed mass function, the resulting constraints on cosmology are largely unchanged, likely due to slight shifts in the astrophysical nuisance parameters that account for the changes in the mass function. In Fig.~\ref{fig:HMF_mass_func}, we show the HMF given each multiplicity function, for both a CDM cosmology and one with 10\% axions and $m_\mathrm{ax} = 10^{-24}$ eV. In the lower panel, we show that axions impact the HMF given each mass function in an almost identical way. Therefore, since other cosmological parameters are not significantly shifted by the choice of mass function, we conclude that the constraints on axion parameters are also unaffected.

\section{Comparison to the observed galaxy main sequence}
\label{sec:AppMS}

\rev{Our model connecting halo mass and galaxy UV luminosity (described in Sec.~\ref{sec:UVLF_compute}) involves a number of intermediate connections: from halo mass to mean stellar mass in Eq.~\eqref{eq:double_power}, to mean star formation rate in Eq.~\eqref{eq:dynamic}, to mean UV luminosity in Eq.~\eqref{eq:UV_lum}. These connections are varied by nuisance parameters over which we marginalize (Table~\ref{Tab:param_priors}). Here, we confirm that our model reproduces observed connections between these intermediate variables, without having seen these additional data.

E.g., the galaxy main sequence is the relation between galaxy stellar mass and star formation rate (SFR). \cite{Popesso2023} fits a parametric relation to a compendium of galaxy main sequence data for \(0 < z < 6\). Fig.~\ref{fig:cMS_check} compares this fit to an estimate of the galaxy main sequence given our UVLF model and HST data. In Eq.~\ref{eq:dynamic}, we assume a linear relation between stellar mass and SFR parameterized by a dynamical time $t_\star$ over which we marginalize, with a redshift dependence given by the Hubble parameter $H(z)$. Since $t_\star$ is degenerate with the amplitude of the stellar mass to halo mass relation $\epsilon_\star$ [Eq.~\eqref{eq:double_power}], we combine these two variables into one, parameterize the combination with a redshift-dependant power-law relation in Eq.~\eqref{eq:variables_end} and then marginalize over the slope $\epsilon_\star^\mathrm{s}$ and intercept $\epsilon_\star^\mathrm{i}$ of this relation. We give the prior and posterior on the slope and intercept in Table~\ref{Tab:param_priors}.

 \begin{figure*}[htbp!]
    \centering
    \includegraphics[trim={2cm 0cm 3cm 0.9cm }, clip,width=0.85\linewidth]{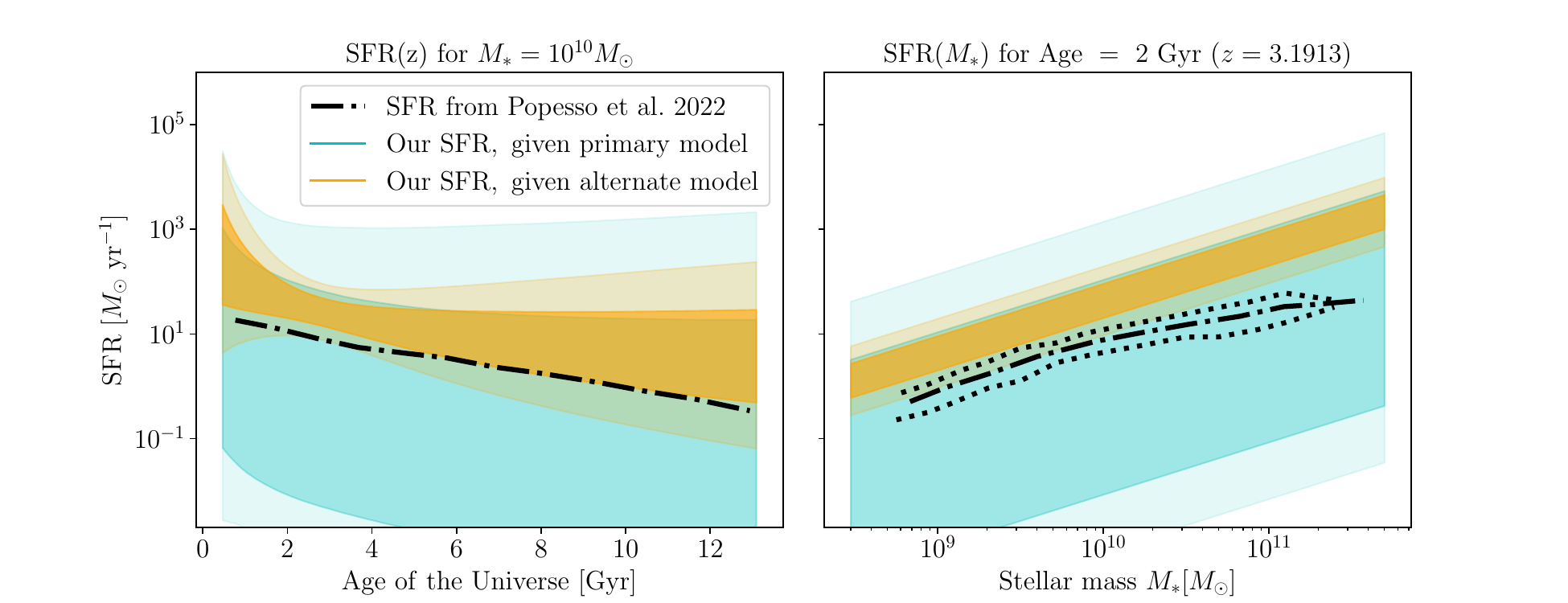}
    \caption{\rev{The mean galaxy star formation rate (SFR) $\overline{\dot{M}_\star}$ as a function of time (left panel) and galaxy stellar mass $M_\star$ (right panel). The left panel fixes $M_\star = 10^{10} M_\odot$, while the right panel fixes age of the Universe \(=\) 2 Gyr (i.e., $z = 3.1913$ for a \textit{Planck} best-fit cosmology). The black dot-dashed line is the measurement from \cite{Popesso2023} (their Fig.~1 for the left panel and their Fig.~2 for the right; both a parametric fit to a compendium of data). The dotted black lines on the right panel indicate the 3$\sigma$ uncertainty from \cite{Popesso2023} (this is not the \textit{scatter} in the SFR, but rather uncertainty in the mean). As described in Appendix~\ref{sec:AppMS}, the colored bands indicate estimates of the galaxy main sequence given our UVLF model and HST data, i.e., independent of the measurements from \cite{Popesso2023}. The cyan bands show an estimate of the posterior on the star formation rate from the primary UVLF model; the orange bands show an estimate of the posterior given the alternative UVLF model presented in Appendix~\ref{sec:AppB} (see text for more details on how these distributions are constructed). The darker and lighter shaded regions respectively indicate the \(68 \%\) and \(95 \%\) credible regions.}} %HW MORE HERE
    %CHANGE LANGUAGE: CYAN IS GIVEN PRIMARY MODEL, RED IS GIVEN ALTERNATIVE MODEL
    \label{fig:cMS_check}
\end{figure*}

In order to examine the relationship that our UVLF model assumes between stellar mass and SFR, we must first disentangle $t_\star$ and $\epsilon_\star$. We do this in two ways. The first, more conservative, approach is to take draws from the posterior on the redshift slope and intercept of the combined $\epsilon_\star/t_\star$ parameter and then divide these by reasonable estimates for \(\epsilon_\star\). For the latter, we take these to be random draws from the uniform prior on the slope and intercept of $\epsilon_\star$. The second approach is to use the alternative UVLF model described in Appendix~\ref{sec:AppB}, where the stellar mass is directly related to the UV luminosity through an empirical relation. We can then take draws from the posterior on the redshift slope and intercept of $\epsilon_\star$ without any explicit sensitivity to \(t_\star\). We then use these draws to normalise the first set of draws described in the first approach. Both approaches are necessarily approximate since, in both UVLF models, we do not fit the galaxy main sequence to any data. The cyan bands in Fig.~\ref{fig:cMS_check} are the 68\% and 95\% limits on the galaxy main sequence given the first approach (based solely on the primary UVLF model); the orange bands are the 68\% and 95\% limits given the second approach (using the alternative UVLF model). The black dot-dashed lines are the observed galaxy main sequence as estimated by \citet{Popesso2023}, from a compendium of previous studies, with the uncertainty on the mean relation shown by the dotted black lines.\footnote{\rev{The orange bands overestimate the star formation rate relative to \citet{Popesso2023}. Since the orange bands are derived from comparing the primary and alternative models of UV luminosity discussed in Appendix~\ref{sec:AppB}, this could be a result of the alternative model preferring lower values of $\epsilon_\star$ relative to the primary model. This result is consistent with the alternative model preferring higher values of $n_\mathrm{s}$ relative to the primary model (and also relative to \textit{Planck} best-fit cosmology), as seen in Fig.~\ref{fig:corner_models}. Higher values of $n_\mathrm{s}$ and lower values of $\epsilon_\star$ could still give a similar UVLF on small scales, but it gives a higher inferred star formation rate when compared to the primary model.}}

The star formation rate from \citet{Popesso2023} shows a decrease at higher stellar mass which is not captured by our model. However, we do not anticipate any significant effect on our results for two reasons. First, the deviation from the linear relation between SFR and stellar mass is increasingly less strong at early times and it is at these times that our analysis occurs (the observed relation that we show is in fact at a slightly lower redshift than the lowest redshift in our data). Second, the suppression in the star formation rate at high stellar mass will manifest in the UVLF as a reduction in the number density at high luminosity. We already, in fact, capture such an effect by the double power law form of the stellar to halo mass relation (where we allow the stellar mass of massive halos to decrease). Since the only observable we use is the UVLF, any additional mass dependence in the star formation rate (or, indeed, in the UV luminosity scatter) is degenerate with the stellar to halo mass relation that we use. Conversely, we anticipate that we can better constrain the UVLF model by using additional information such as the observed galaxy main sequence.
}

\bibliography{main_bibliography.bib}

\end{document}